\documentclass[%
preprint, superscriptaddress, 12pt,superscriptaddress,
 amsmath,amssymb,
 aps,
 showkeys,
]{revtex4-1}

\usepackage{scalerel}
\usepackage{siunitx}
\usepackage{dcolumn} 
\usepackage{titlesec}

\titlespacing*{\section}
{0pt}{1\baselineskip}{.1\baselineskip}

\titlespacing*{\subsection}
{0pt}{4.5ex plus 1ex minus .2ex}{.5\baselineskip}
\titleformat{\section}
  {\centering\bfseries\MakeUppercase}{\thesection.}{1em}{}
\titleformat{\subsection}
  {\centering\bfseries\MakeUppercase\small}{\thesubsection.}{1em}{}

\usepackage{xstring}
\usepackage{tikz}

\usepackage{fancyhdr} 
\usepackage{graphicx}
\usepackage{dcolumn}
\usepackage{float}

\usepackage{mathtools}
\usepackage{bm}

\newcommand{\SM}{Supporting Information}
\usepackage{xcolor}  
\usepackage{soul}  
\usepackage{natbib}
\usepackage[english]{babel}
\usepackage[utf8]{inputenc}
\usepackage{url}
\usepackage[colorlinks = true,
            linkcolor = red,
            urlcolor  = blue,
            citecolor = red,
            anchorcolor = black]{hyperref}

 \usepackage[justification=raggedright,singlelinecheck=false,labelfont=bf,labelsep=period]{caption,subcaption}
\usepackage{ulem}
\makeatletter
\renewcommand\hl{%
  \bgroup
  \UL@protected\def\sout{\bgroup \ULdepth =-.8ex \ULset}%
  \markoverwith{\textcolor{yellow}{\rule[-.5ex]{.1pt}{2.5ex}}}%
  \ULon}
\makeatother

\usepackage{scalerel}
\usepackage{tikz}
\usetikzlibrary{svg.path}
\definecolor{orcidlogocol}{HTML}{A6CE39}
\tikzset{
  orcidlogo/.pic={
    \fill[orcidlogocol] svg{M256,128c0,70.7-57.3,128-128,128C57.3,256,0,198.7,0,128C0,57.3,57.3,0,128,0C198.7,0,256,57.3,256,128z};
    \fill[white] svg{M86.3,186.2H70.9V79.1h15.4v48.4V186.2z}
                 svg{M108.9,79.1h41.6c39.6,0,57,28.3,57,53.6c0,27.5-21.5,53.6-56.8,53.6h-41.8V79.1z M124.3,172.4h24.5c34.9,0,42.9-26.5,42.9-39.7c0-21.5-13.7-39.7-43.7-39.7h-23.7V172.4z}
                 svg{M88.7,56.8c0,5.5-4.5,10.1-10.1,10.1c-5.6,0-10.1-4.6-10.1-10.1c0-5.6,4.5-10.1,10.1-10.1C84.2,46.7,88.7,51.3,88.7,56.8z};}}
\newcommand\orcidicon[1]{\href{https://orcid.org/#1}{\mbox{\scalerel*{
\begin{tikzpicture}[yscale=-1,transform shape]
\pic{orcidlogo};
\end{tikzpicture}
}{|}}}}

\usepackage{etoolbox}

\makeatletter

\begin{document}

\title{Mapping the Pathways of Photo-induced Ion Migration in Organic-inorganic Hybrid Halide Perovskites}

\author{Taeyong Kim}
\affiliation{Department of Mechanical Engineering, University of California, Santa Barbara, CA 93106, USA}
\affiliation{Department of Mechanical Engineering,
Seoul National University, Seoul, 08826, Republic of Korea}

\author{Soyeon Park}
\affiliation{National Renewable Energy Laboratory, Golden, CO 80401, USA}

\author{Vasudevan Iyer}
\affiliation{Center for Nanophase Materials Sciences, Oak Ridge National Laboratory, Oak Ridge, TN 37830, USA}

\author{Qi Jiang}
\affiliation{National Renewable Energy Laboratory, Golden, CO 80401, USA}

\author{Usama Choudhry}
\affiliation{Department of Mechanical Engineering, University of California, Santa Barbara, CA 93106, USA}

\author{Gage Eichman}
\affiliation{Center for Nanophase Materials Sciences, Oak Ridge National Laboratory, Oak Ridge, TN 37830, USA}

\author{Ryan Gnabasik}
\affiliation{Department of Mechanical Engineering, University of California, Santa Barbara, CA 93106, USA}

\author{Benjamin Lawrie}
\email{lawriebj@ornl.gov}
\affiliation{Center for Nanophase Materials Sciences, Oak Ridge National Laboratory, Oak Ridge, TN 37830, USA}
\affiliation{Materials Science and Technology Division, Oak Ridge National Laboratory, Oak Ridge, TN 37830, USA}

\author{Kai Zhu}
\email{kai.zhu@nrel.gov }
\affiliation{National Renewable Energy Laboratory, Golden, CO 80401, USA}

\author{Bolin Liao}
\email{bliao@ucsb.edu} \affiliation{Department of Mechanical Engineering, University of California, Santa Barbara, CA 93106, USA}

\date{\today}

\begin{abstract}
Organic-inorganic hybrid perovskites (OIHPs) exhibiting exceptional photovoltaic and optoelectronic properties are of fundamental and practical interest, owing to their tunability and low manufacturing cost. For practical applications, however, challenges such as material instability and the photocurrent hysteresis occurring in perovskite solar cells under light exposure need to be understood and addressed. While extensive investigations have suggested that ion migration is a plausible origin of these detrimental effects, detailed understanding of the ion migration pathways remains elusive. Here, we report the characterization of photo-induced ion migration in OIHPs using \textit{in situ} laser illumination inside a scanning electron microscope, coupled with secondary electron imaging, energy-dispersive X-ray spectroscopy and cathodoluminescence with varying primary electron energies. Using methylammonium lead iodide (MAPbI$_3$), formamidinium lead iodide (FAPbI$_3$) and hybrid formamidinium-methylammonium lead iodide as model systems, we observed photo-induced long-range migration of halide ions over hundreds of micrometers and elucidated the transport pathways of various ions both near the surface and inside the bulk of the OIHPs, including a surprising finding of the vertical migration of lead ions. Our study provides insights into ion migration processes in OIHPs that can aid OIHP material design and processing in future applications.  

\end{abstract}

\maketitle

\section{Introduction}
Rapid progress has been made in the past decade in improving the power conversion efficiency of photovoltaic cells based on organic-inorganic hybrid perovskites (OIHPs). Recent reports have described power conversion efficiencies exceeding 25\%~\cite{jung2019efficient,Qi_Nature_2022,kim2020high,Soyeon_Review}. The remarkable performance has been attributed to OIHPs' unique optoelectronic properties such as a direct band gap and sharp absorption edge near visible wavelengths~\cite{Baikie_JMCA_2013,Stranks_Natnano_2015,yin2015superior}, large photocarrier diffusion lengths \cite{Shi_Science_2015,guo2017long} and high tolerance to defects and grain boundaries~\cite{ball2016defects,brenes2017metal}, which are key factors for photovoltaic cells~\cite{Kojima_JACS_2009, Grancini_NatMat_2019,Green_SCETable_2021}. Additionally, OIHPs have shown other interesting properties, such as their tunability~\cite{Kulkarni_JMCA_2014}, moderate mobilities of charge carriers ~\cite{Shi_Science_2015} and ions~\cite{Bakulin_JPCL_2015}, strong polaronic coupling~\cite{Kandada_JPCL_2020, Grancini_NatMat_2019}, and a prominent photostriction effect \cite{zhou_ncomm_2016}. Coupled with their cost-effective low-temperature solution processability~\cite{song2017technoeconomic}, these unusual optoelectronic properties suggest wide-ranging practical optoelectronic applications of OIHPs beyond solar cells. 

In addition to investigations of device efficiency optimization, extensive prior studies have focused on understanding and improving the stability of OIHPs~\cite{Boyd_CR_2019}, which is a crucial limiting factor for long-term practical applications. It has been well established experimentally that many extrinsic factors can lead to degradation in these hybrid perovskite materials~\cite{Bella_Science_2016,Boyd_CR_2019}, including exposure to moisture ~\cite{Bella_Science_2016, Christians_JACS_2015} and oxygen~\cite{Bryant_EES_2016}, thermal decomposition~\cite{Conings_AEM_2015}, and light-induced chemical reactions~\cite{Domanski_EES_2017,Burschka_Nature_2013}. Corresponding strategies have been developed to counter some of the degradation factors, e.g. OIHP encapsulation to avoid exposure to moisture or oxygen~\cite{lee2018low,cheacharoen2018design}.

Many of the instability factors are related to the weak chemical bonds~\cite{ma2019supercompliant} and strongly anharmonic lattice dynamics~\cite{yaffe2017local} in OIHPs that also lead to ultralow thermal conductivity~\cite{haque2020halide} and facile ion migration induced by temperature~\cite{lai2018intrinsic}, electric field~\cite{Xiao_NatMat_2015}, or light~\cite{Kim_NatMat_2018}. Since temperature gradients, electric fields, and light exposure are unavoidable in the normal operation of solar cells, it is recognized that ion migration processes play an essential role in OIHP-based devices~\cite{Yuan_ACR_2016,bi2021mitigating}. In this light, recent studies have observed ion-migration occurring in OIHPs and its impact on the material morphology~\cite{Xiao_NatMat_2015}, composition and device stability~\cite{Burschka_Nature_2013,Snaith_JPCL_2014,Yuan_ACR_2016}. In addition, these studies have reported that the ion migration might cause the abnormal photocurrent hysteresis effect, which, in turn, affects fill factors and solar cell device performance~\cite{Snaith_JPCL_2014}. Unlike extrinsic factors such as moisture and oxygen, ion migration is intrinsic in OIHPs, likely linked to certain strongly anharmonic lattice vibration modes involving the mobile ions~\cite{mizusaki1983ionic}, and, therefore, requires more sophisticated mitigation strategies~\cite{bi2021mitigating}. On the other hand, ion migration represents a possible mechanism for the large static dielectric constant in OIHPs~\cite{lin2015electro} that can effectively screen charged defects and benefit charge transport. Some studies have also suggested that the redistribution of ions as a result of migration can lead to optimized p-i-n structures~\cite{deng2015light} or healed charge trapping centers~\cite{ghosh2017ion} and improve the device performance. Besides, efficient ion migration in OIHPs also implies potential applications of these materials as ionic conductors~\cite{Kim_NatMat_2018}. These examples highlight the importance of understanding the ion migration processes in OIHPs~\cite{Yuan_ACR_2016}. 

Previous works have reported light-induced migration of ions such as the halide anion (I$^-$), the metal cation (Pb$^{2+}$), and the organic cations (MA$^+$ = Methylammonium, FA$^+$ = Formamidinium) \cite{Kim_NatMat_2018, Zhao2017,shirzadi2022deconvolution,barbe2018localized}. In parallel, prior computational studies have shown that the migration rate is determined by the activation energy and that I$^-$ is more likely to migrate  owing to its lowest activation energy among other constituent ions (I$^-$: 0.08 - 0.58 eV in Refs.~\cite{Azpiroz_EES_2015,Eames_NatComms_2015,Haruyama_JACS_2015, futscher2019quantification}; MA$^{+}$: 0.46 - 0.84 eV in Refs.~\cite{Azpiroz_EES_2015,Eames_NatComms_2015,Haruyama_JACS_2015,futscher2019quantification};~Pb$^{2+}$: 0.8 - 2.31 eV in Refs.~\cite{Azpiroz_EES_2015,Eames_NatComms_2015}). Experimental measurements have corroborated this, with the diffusivity of I$^-$ ($\sim10^{-12}$ to $\sim10^{-9}$~cm$^2$~s$^{-1}$~\cite{Azpiroz_EES_2015,McGovern_ACSAEM_2021,futscher2019quantification}) far exceeding that of organic cations (MA: $\sim10^{-16}$ to $\sim10^{-11}$~cm$^2$~s$^{-1}$~\cite{Azpiroz_EES_2015,McGovern_ACSAEM_2021,futscher2019quantification}). In most previous experimental studies, the Pb$^{2+}$ ion was assumed to be immobile up to the available detection limits~\cite{yuan2016electric}. Despite the abundant previous literature on photo-induced ion migration in OIHPs, the detailed microscopic migration pathways of various ions remain unclear. In particular, ion migration both parallel to the sample surface (lateral ion migration) and perpendicular to the sample surface (vertical ion migration) can occur given the light intensity variation along the lateral direction and the finite optical absorption depth into the sample. Therefore, experimental tools that are sensitive to the distribution of ions along both the lateral and vertical directions are highly desirable. 

In this work, we focus on photo-induced ion migration in OIHPs. We incorporated a laser source \textit{in situ} into a scanning electron microscope (SEM) and combined secondary electron imaging (SEI), energy-dispersive X-ray spectroscopy (EDS) and cathodoluminescence (CL) with different primary electron (PE) energies to directly map the three-dimensional (3D) ion migration pathways in two archetypal OIHPs, MAPbI$_3$, FAPbI$_3$ and their hybrid. We demonstrated that both lateral and vertical migrations of various ion species driven by light exposure can be mapped in this manner, which further correlate to the optoelectronic properties of OIHPs. This information is critical for designing more stable and efficient devices based on OIHPs.

\section{Results}
\subsection{Probing ion distribution within different depths}
\begin{figure}
{
\includegraphics[width=.75\textwidth,keepaspectratio]{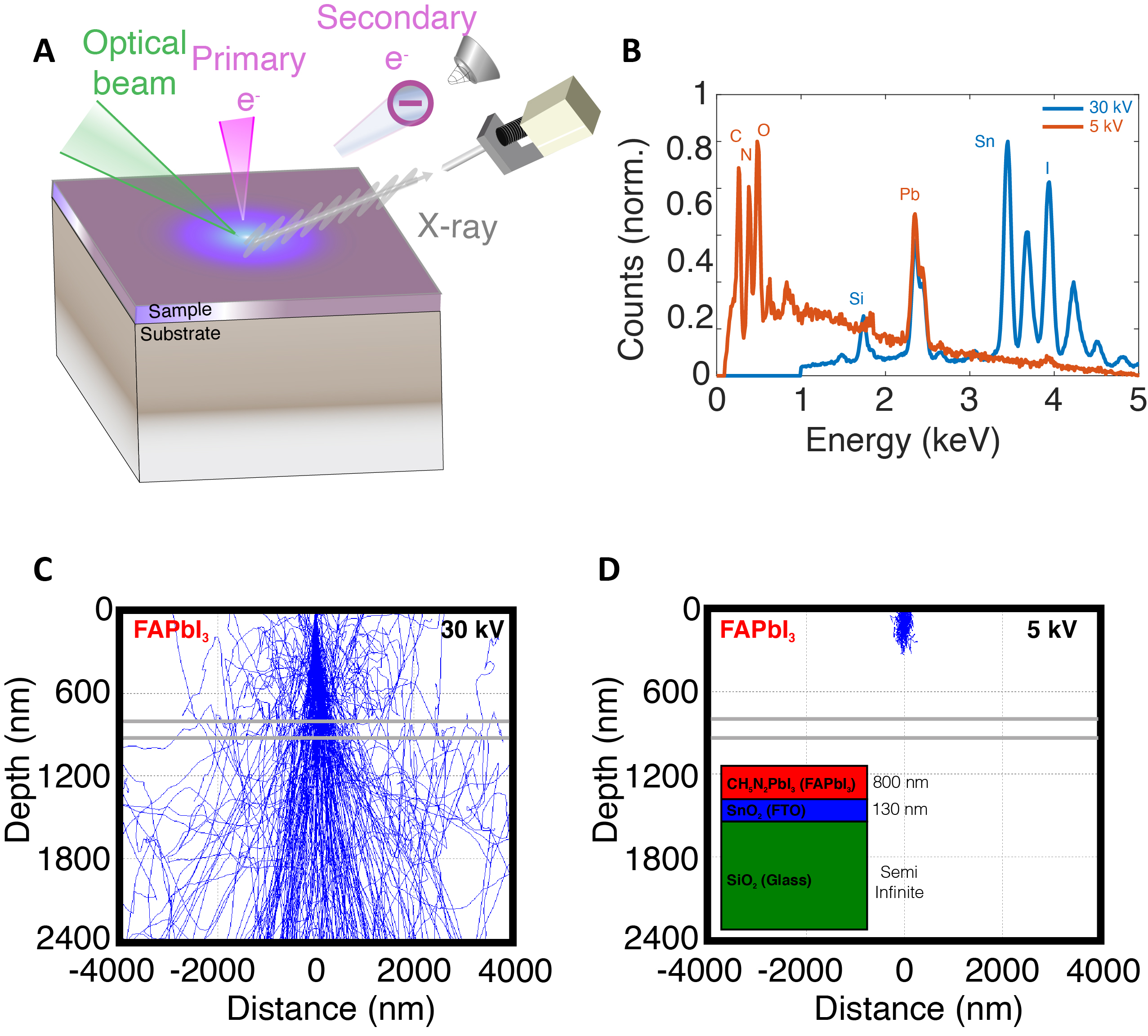}
\caption{\label{fig:Schematic}
\textbf{Schematic illustration of the experimental setup and sample structure.} (A) Illustration of the SEM experimental platform incorporating an \textit{in situ} laser beam with SEI and EDS detectors. (B) The measured EDS spectra of FAPbI$_3$ without light exposure at the primary electron energies of 30 keV (blue line) and 5 keV (red line). The corresponding elements are annotated near each characteristic peak. Also shown are the simulated trajectories of primary electrons with energies of (C) 30 keV, and (D) 5 keV. While the primary electrons can penetrate the entire sample stack at 30 keV, those at 5 keV are mostly confined within 200 nm near the FAPbI$_3$ surface. The interfaces between FAPbI$_3$ and fluorine doped tin oxide (FTO), and between FTO and the glass substrate are labeled with gray lines. The inset in (D) shows the structure of the FAPbI$_3$ sample stack.}
}
\end{figure}

Our experimental setup for the SEI and EDS measurements is hosted at the University of California Santa Barbara (UCSB) and is identical to that described in Ref.~\cite{Kim_NL}, and the principle of operation is schematically illustrated in Fig.~\ref{fig:Schematic}A. An optical beam from a fiber laser (Clark-MXR IMPULSE, photon energy: $\sim 2.4$ eV; repetition rate: 5 MHz) was coupled into the SEM sample chamber and focused onto the sample, with a beam diameter of $50 ~\mu$m. To examine the impact of the photo-induced ion migration, the optical power and the exposure time were varied, ranging from $\sim2.5$ to $\sim6.7$ mW (optical fluence: $\sim 39.8 - 106.6$ $\mu$J~cm$^{-2}$), and from 5 to 180 minutes, respectively. These optical parameters for the SEI and EDS measurements were chosen to minimize the laser-induced heating effect, and we estimated that the highest laser-induced temperature rise in our experiment was within 30 K (see \SM~Sec.~V). Then, changes in the secondary electron (SE) emission and the elemental composition induced by the optical exposure were mapped \textit{in situ} with concurrent SEI and EDS mapping. The depth-dependent ion distribution was probed by comparing the response to PEs with different kinetic energies (30 keV and 5 keV) and different penetration depths into the sample. The CL measurements coupled with \textit{in situ} light-exposure were conducted at the Center for Nanophase Materials Sciences (CNMS) at the Oak Ridge National Laboratory (ORNL), and the corresponding experimental details are given in Methods. The procedure for sample fabrication is also described in Methods. The structure of the sample stack is illustrated in the inset of Fig.~\ref{fig:Schematic}D. A high vacuum of $1 \times 10^{-6}$ torr was maintained inside the SEM chambers for all measurements reported here, thus minimizing the influence of moisture or oxygen. Representative EDS spectra of a pristine FAPbI$_3$ sample deposited on a fluorine-tin-oxide (FTO) coated glass are given in Fig.~\ref{fig:Schematic}B (the MAPbI$_3$ spectra is given in the \SM~Sec.~I). In the EDS spectra, peaks corresponding to characteristic X-ray lines of various chemical elements can be seen. The intensity of these characteristic peaks also depends on the PE kinetic energy used. With 30-keV PEs, peaks near 1.74 keV, 2.35 keV, 3.44 keV, and 3.94 keV were prominent, which were characteristic X-ray emissions from silicon (K$_\alpha$ band), lead (M$_\alpha$ band), tin (L$_\alpha$ band) and iodine (L$_{\alpha1}$ band)~\cite{Mandal_JMS_2020,Cheng_JMS_2016,Maity_SciRep_2018, Lin_EDS_2011}. Lead and iodine were from the FAPbI$_3$ layer, while tin and silicon only existed in the substrate, signaling a deep penetration of the 30-keV PEs. In contrast, with 5-keV PEs, we found that the tin and iodine bands were suppressed, the lead band remained, while other lower-energy peaks emerged near 277 eV, 392 eV, and 525 eV, corresponding to the characteristic X-ray emissions from carbon, nitrogen and oxygen~\cite{Goldsteinbook}. Additional EDS spectra with intermediate PE energies (10 keV and 20 keV) are shown in the \SM~Sec.~I. 

To quantitatively understand the probing depth of PEs with different kinetic energies, we conducted Monte Carlo simulations of the PE trajectories with 30-keV and 5-keV kinetic energies and the results were shown in Fig.~\ref{fig:Schematic}C and D. The detailed parameters used in our simulation were described in the \SM~Sec.~II. As shown in Fig.~\ref{fig:Schematic}C and D, many of the 30-keV PEs penetrated the entire FAPbI$_3$ layer and into the substrate, while those at 5 keV were mostly confined within 200 nm near the surface of FAPbI$_3$. This indicates that the EDS spectra obtained with PEs with different kinetic energies are sensitive to ion distributions at different depths into the sample. Additionally, we note here that the self-absorption depth of the X-ray photons emitted inside the samples was comparable to or greater than the sample thickness, except for those X-ray photons emitted from carbon, nitrogen and oxygen (see the \SM~Sec.~IV for further details), indicating that our measurement was sensitive to the PE penetration depth rather than the X-ray absorption depth except for carbon, nitrogen and oxygen.

\begin{figure}
{
\includegraphics[width=0.85\textwidth,height=\textheight/3,keepaspectratio]{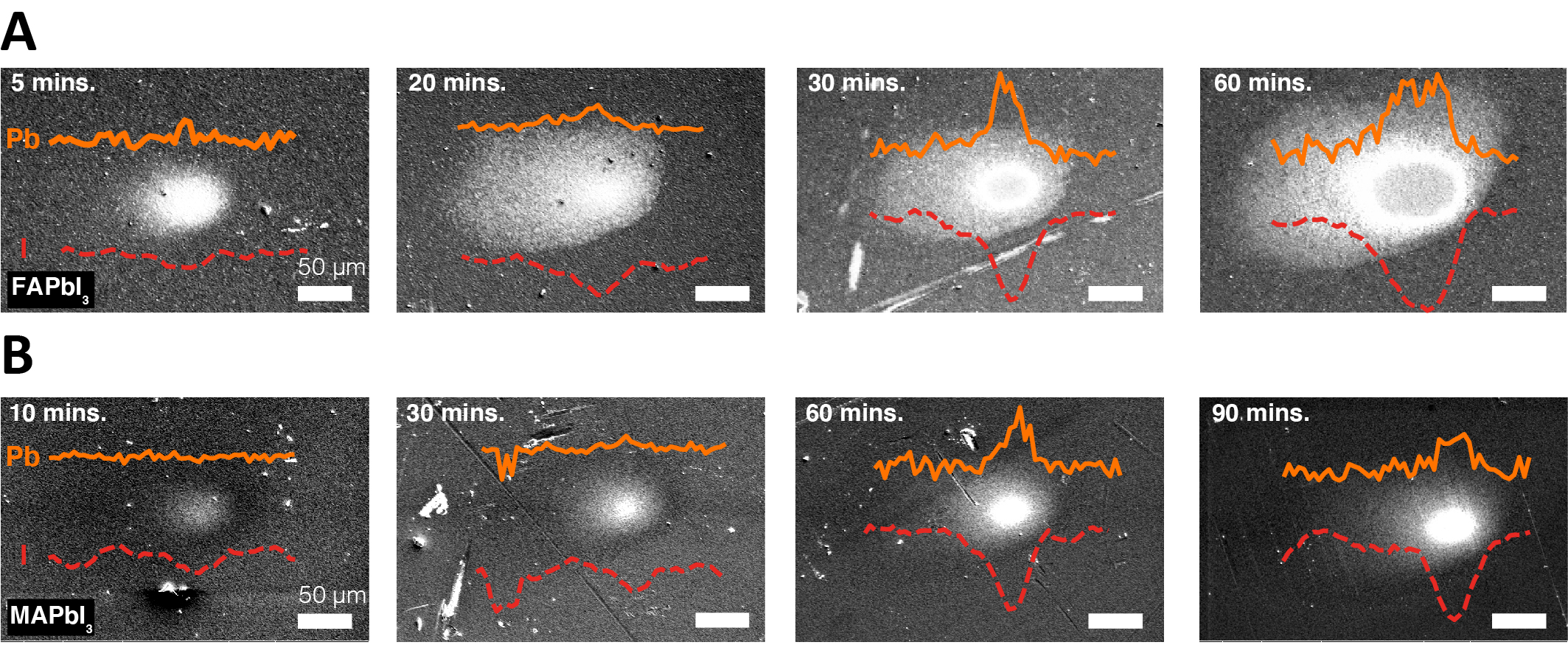}
\caption{\label{fig:SEimages}
\textbf{Secondary electron images with line profiles of iodine and lead distribution after optical beam exposure.} (A) SE images of  FAPbI$_3$ after laser illumination for 5 to 60 minutes. (B) SE images of MAPbI$_3$ after laser illumination for 10 to 90 minutes. The red dashed lines (orange solid lines) indicate the X-ray counts for iodine (lead) at 3.94 keV (2.35 keV) across the illuminated area as measured using EDS. The iodine line was probed by 30-keV PEs and the lead line was probed by 5-keV PEs. For all the data presented here, an optical fluence of $\sim 106.6$ $\mu$J~cm$^{-2}$ was used.} 
}
\end{figure}

Our experimental configuration permitted us to characterize the photo-induced ion movement along with their impact on the SE yield in an \textit{in situ} manner. Figure \ref{fig:SEimages} shows several SE images of FAPbI$_3$ and MAPbI$_3$ samples, which were taken after the laser beam exposure (optical fluence: $106.6$ $\mu$J~cm$^{-2}$ for both FAPbI$_3$ and MAPbI$_3$) at the center of the field of view for variable laser dwell times. As shown in Fig.~\ref{fig:SEimages}A, for FAPbI$_3$, the SE image contrast became brighter after the laser exposure for 5 minutes. Further exposure of up to 20 minutes led to the bright contrast expanding beyond the laser spot size. Under prolonged optical exposure of $\gtrsim 30$ minutes, an area with dark contrast emerged near the center of the beam area, forming a ring-shaped profile that was surrounded by bright contrast at the outer region. We also observed that additional optical exposure caused further expansion of the bright ring, while the surrounding bright contrast area far exceeded the size of the optical beam. In contrast, the SE images of MAPbI$_3$ after light exposure exhibited a slower temporal evolution of the bright contrast, likely due to a larger bandgap in MAPbI$_3$ and, thus, a weaker optical absorption. The size of the area with bright contrast after longer exposure times was observed to be comparable to the optical beam size. At lower optical fluences, qualitatively similar contrast in SE images was seen, although the required exposure time was much longer (see the \SM~Sec.~I for additional data). We can further correlate the observed bright SE contrast with the ion distribution measured \textit{in situ} by EDS. In particular, the X-ray counts corresponding to the characteristic iodine line at 3.94 keV (probed by 30-keV PEs) and lead line at 2.35 keV (probed by 5-keV PEs) were measured along a horizontal line across the center of the area exposed to the optical beam and were given as red dashed lines in Fig.~\ref{fig:SEimages}A and B. In general, we observed a decrease of iodine counts within the area exposed to light, and the area with an iodine deficiency increased with prolonged light exposure. Particularly in FAPbI$_3$ (Fig.~\ref{fig:SEimages}A), the iodine concentration profile agreed well with the spatial extent of the bright contrast, suggesting long-ranged migration of iodine ions far outside the illuminated region. However, the iodine concentration profile is less consistent with the bright ring contrast observed in FAPbI$_3$. In contrast, although the lead concentration profile did not correlate to the overall spatial range of the bright contrast, a two-peak structure developed in FAPbI$_3$ after 30-minute light exposure that matches the bright ring-shaped contrast, indicating a lead-rich phase was responsible for the significantly enhanced SE emission from the ring-shaped area. With our current technique, however, the precise composition of this lead-rich phase could not be determined, since the 30-keV PEs probed the iodine concentration inside the entire thickness of the sample, not only at the surface.

\subsection{Quantification of ion migration processes}
\begin{figure}
{
\includegraphics[width=\textwidth,height=\textheight/2,keepaspectratio]{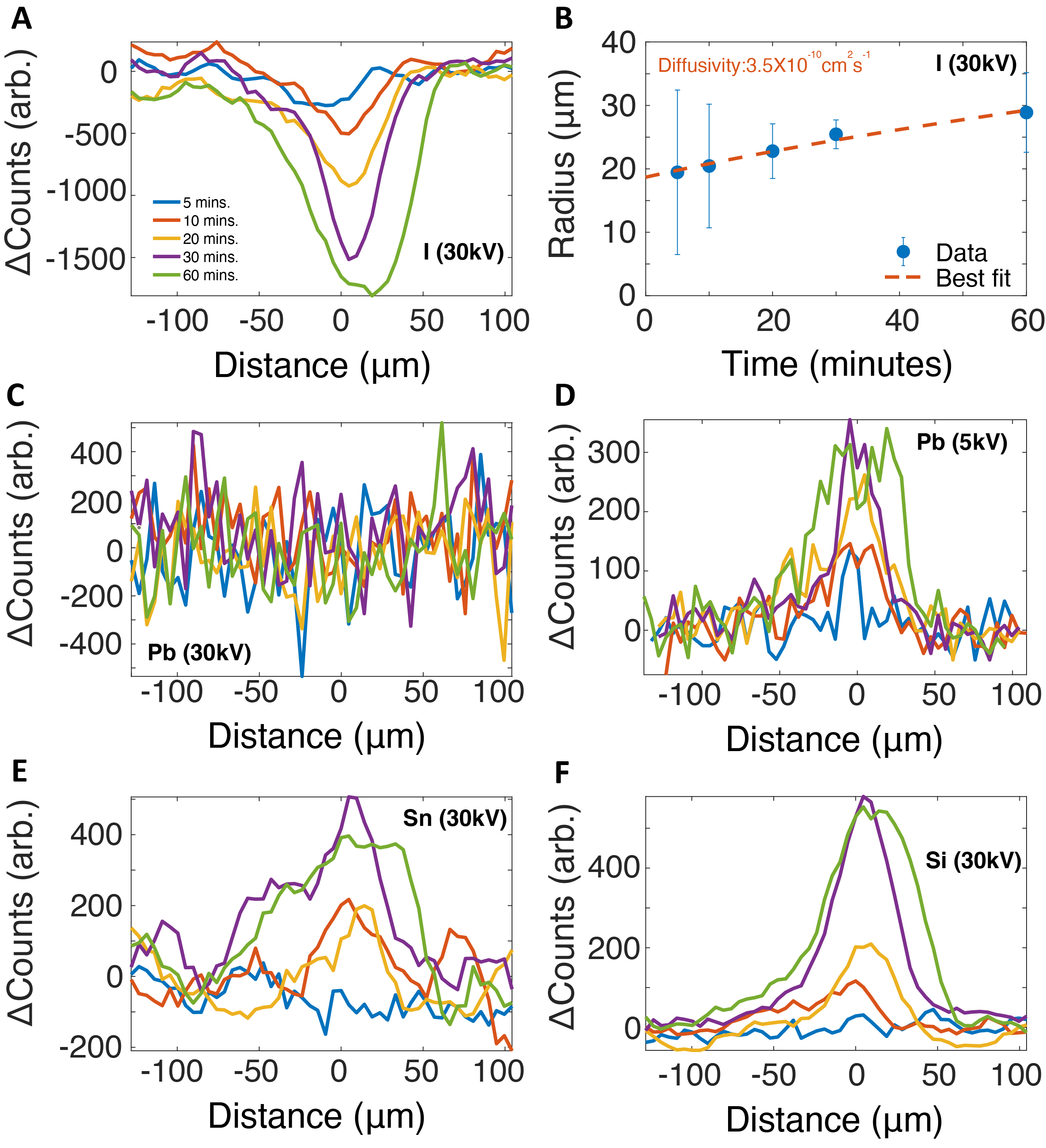}
\caption{\label{fig:Lineprof}
\textbf{EDS elemental analysis of FAPbI$_3$ after optical exposure.} An optical fluence of $\sim 106.6$ $\mu$J~cm$^{-2}$ was used for this dataset. (A) The change in the X-ray counts corresponding to iodine as probed by 30-keV PEs. The change in counts was calculated relative to the value measured from pristine areas without light exposure. (B) The corresponding fitted radii of the iodine deficiency distributions versus the exposure times measured by 30-keV PEs. The error bar indicates numerically determined 95\% confidence intervals. Also shown are the measured EDS linescan datasets for (C) lead measured by 30-keV PEs, (D) lead measured by 5-keV PEs, (E) tin measured by 30-keV PEs, and (F) silicon measured by 30-keV PEs. } 
}
\end{figure}

We can quantify the ion migration processes by carefully examining the ion distributions mapped by EDS after different light exposure times. We first focus on the iodine ions, which are known to be the most mobile species in OIHPs. In Fig.~\ref{fig:Lineprof}A, we show the changes in the characteristic iodine X-ray photons at 3.94 keV from FAPbI$_3$ near the illuminated region. The changes in the counts were calculated relative to the value measured in the pristine region without light exposure. The changes in counts were measured using EDS with 30-keV PEs and plotted as a function of the distance from the center of the optical beam and after different light exposure durations. A reduction of the iodine X-ray counts was observed in the illuminated region, signaling a deficiency of iodine as a result of light exposure. Furthermore, the iodine deficiency kept increasing and its spatial distribution kept expanding as a function of exposure time. Since the 30-keV PEs probed the entire sample thickness, the observed iodine deficiency indicated the amount of iodine inside the sample was reduced, likely due to the decomposition of FAPbI$_3$ and the formation of I$_2$ vapor escaping to the vacuum~\cite{albrecht1977kinetics}. In addition, the expanding spatial profile of the iodine deficiency signals migration of iodine ions over a lengthscale of a few hundred micrometers, significantly exceeding the optical beam size, driven by the iodine concentration gradient due to the I$_2$ escaping from the surface. From the time dependence of the spatial profile of the iodine deficiency, we can estimate the diffusivity of iodine ions, as shown in Fig.~\ref{fig:Lineprof}B. The extracted iodine ion diffusivity is around $3.5 \times 10^{-10}$ cm$^2$~s$^{-1}$, which is within the range of previously reported values~\cite{Azpiroz_EES_2015,McGovern_ACSAEM_2021,futscher2019quantification}. A detailed analysis of the diffusion process was described in the \SM~Sec.~III.  

Next, we focus on the lead ions, whose distribution can be mapped by PEs with both 30-keV and 5-keV kinetic energies. As shown in Fig.~\ref{fig:Lineprof}C and D, the lead ion distribution probed by 30-keV PEs remained almost unchanged as a result of light exposure, while its distribution probed by 5-keV PEs showed a significant increase with light exposure. Since the 30-keV PEs can probe the entire sample thickness and the 5-keV PEs only probe near the sample surface, this surprising finding implied that the migration of lead ions mainly occurred along the vertical direction, namely moving from inside the sample towards the sample surface. More interestingly, the lead distribution near the surface as probed by the 5-keV PEs showed a double-peak structure after a long exposure, consistent with the bright ring contrast observed in the SE images shown in Fig. \ref{fig:SEimages}A. As discussed before, this suggests that a lead-rich phase is responsible for the enhanced SE emission in the bright ring region.    

Another surprising finding was the observed migration of silicon and tin ions from the substrate, as shown in Fig.~\ref{fig:Lineprof}E and F. While a small trace of silicon was detected with the 5-keV PEs (as shown in Fig.~\ref{fig:Schematic}B), the silicon distribution did not change appreciably due to the light exposure (See the \SM~Sec.~I for additional data). However, the silicon and tin distributions as probed by the 30-keV PEs showed a significant increase with prolonged light exposure, suggesting increased silicon and tin concentration deep inside the bulk of the FAPbI$_3$ sample. This can be induced by the isovalent lead vacancies created deep inside the bulk of the FAPbI$_3$ sample after the lead ions migrated towards the sample surface. This observation implies that ion migration can lead to unwanted interactions between the OIHPs and the charge transport layers that can be detrimental to solar cells based on OIHPs~\cite{li2019suppressing}. Lastly, we also observed a decrease of nitrogen, carbon and oxygen concentration near the sample surface as probed by 5-keV PEs due to light exposure (See the \SM~Sec.~I for additional data). However, we could not quantify the diffusivity of the organic ions due to the relatively low EDS sensitivity and, thus, a low signal-noise ratio.   

\begin{figure}
{
\includegraphics[width=0.7\textwidth,height=\textheight/3,keepaspectratio]{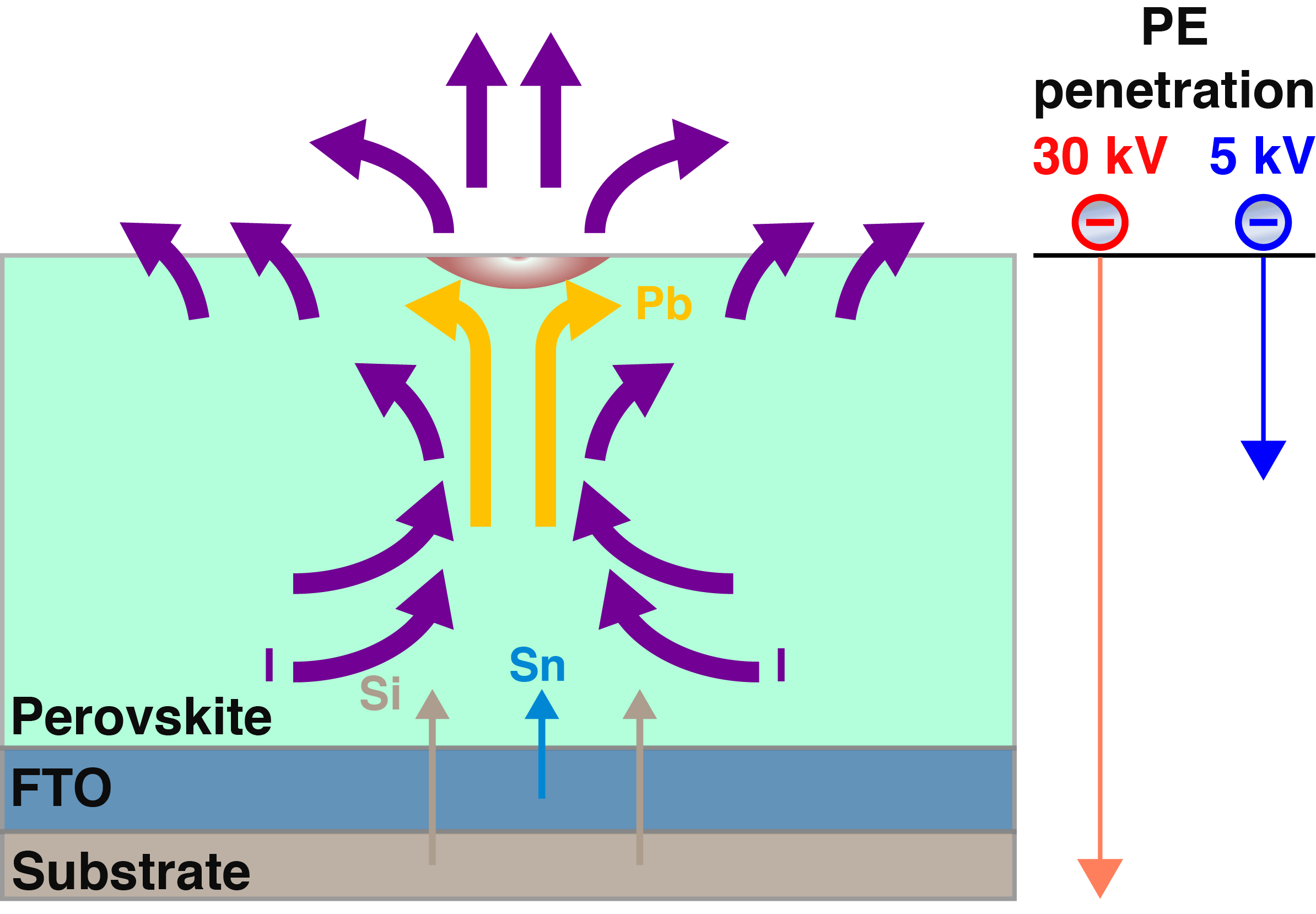}
\caption{\label{fig:picture}
\textbf{Schematic illustration of the photo-induced ion migration pathways in OIHPs.} Iodine ions migrate from the bulk towards the surface due to photo-induced decomposition and I$_2$ escaping to the vacuum. Lead ions migrate vertically toward the surface. At the same time, elements in the substrate such as tin and silicon penetrate across the sample/substrate interface. Right panel indicates schematically the PE penetration depths through the specimen, which depends on the PE kinetic energy.}
}
\end{figure}

Based on the observations described above, we summarize the microscopic picture of photo-induced ion migration in FAPbI$_3$ schematically in Fig.~\ref{fig:picture}. Photo-induced decomposition of FAPbI$_3$ into PbI$_2$ and, further, into I$_2$ vapor leads to an iodine deficiency near the surface, which creates an iodine concentration gradient that drives the diffusion of iodine ions towards the surface. The lateral iodine concentration gradient created by the Gaussian distribution of the laser beam intensity further induces iodine ion migration towards the center of the laser illuminated area. In the meantime, lead ions also migrate from the bulk towards the surface. Since lead cannot escape the sample surface in vapor form, the accumulation of lead near the center of the illuminated area starts to drive lead ions to move outwards near the surface after longer light exposure, creating the double-peak lead concentration profile that corresponds to the bright ring contrast in SE images. As a consequence of lead ion migration, the lead vacancies left behind deep inside the sample bulk induce the migration of isovalent tin and silicon ions from the substrate into the sample, creating new defects and phases that might be detrimental to OIHP-based devices.

\begin{figure}[hbt!]
{
\includegraphics[width=\textwidth,keepaspectratio]{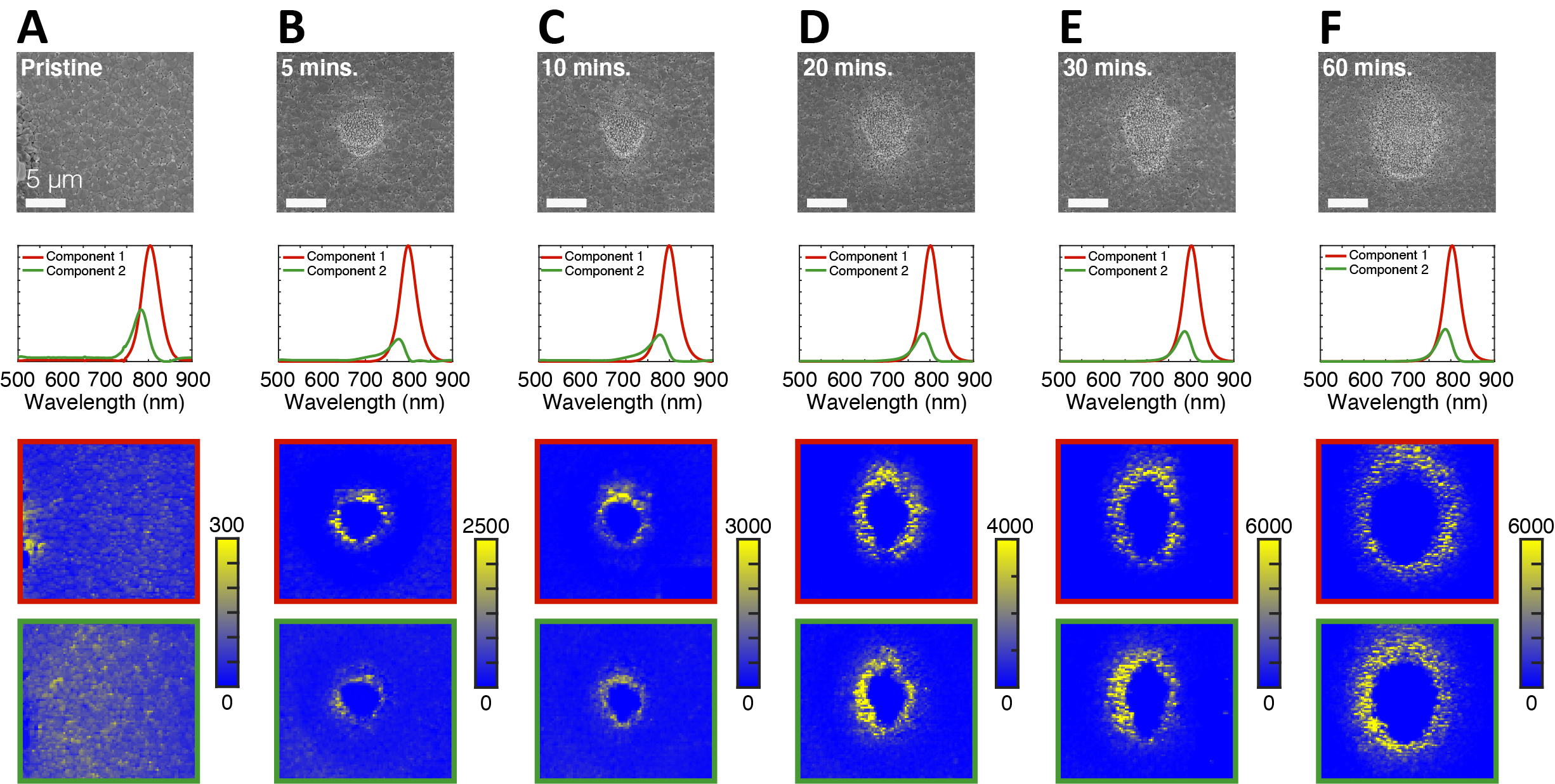}
\caption{\label{fig:CL}
\textbf{Cathodoluminescence analysis of 97\% FAPbI$_3$/3\% MAPbBr$_3$ after \textit{in situ} pulsed laser exposure.} Correlated SE images (first row) and CL NMF decomposition (row 2 to 4) for (a) the pristine sample and the same sample after (B) 5 minutes of optical exposure, (C) 10 minutes of optical exposure, (D) 20 minutes of optical exposure, (E) 30 minutes of optical exposure, and (F) 60 minutes of optical exposure. The two spectral components of each NMF decomposition are illustrated in the second row, and the intensity maps associated with each of those components are illustrated in the third and the fourth row, respectively. The color bars represent the absolute CL counts and highlight the increase in band-edge CL intensity as a function of increased laser exposure time.}
}
\end{figure}

\subsection{Correlating ion migration with cathodoluminescence}
We note here that Barb\'{e} et al. observed a light-induced ring-shaped area with enhanced photoluminescence~\cite{barbe2018localized} in MAPbI$_3$ and they attributed it to the formation of a thin layer ($< 20$ nm) of PbI$_2$ using Raman spectroscopy. However, no photoluminescence near the bandgap of PbI$_2$ ($\sim 520$ nm to 540 nm)~\cite{ferreira1996optical} was observed. To further examine the correlation between light exposure and the optoelectronic properties of OIHPs, we conducted \textit{in situ} CL measurements of a hybrid OIHP, 97\% FAPbI$_3$ with 3\% MAPbBr$_3$. For this measurement, a femtosecond laser with 495 nm wavelength and 1.25 mW average power was focused to a 5 $\mu$m spot (optical fluence: 80 $\mu$J~cm$^{-2}$) on the sample using a parabolic mirror in the SEM. The laser-induced temperature rise in this case was around 90 K due to the small beam size (see~\SM~Sec.~V for details). Despite similar optical fluences, the higher steady-state temperature rise in this case could also induce ion migration thermally. The same parabolic mirror was used to collect CL generated by the perovskite film. Notably, substantial research efforts in recent years have probed ion migration and perovskite decomposition under optical and electron-beam excitation and in the presence of various environmental factors~\cite{shirzadi2022deconvolution,milosavljevic2016low,yuan2016degradation,xiao2015mechanisms}.  However, under the electron-beam conditions used here (5-keV PE and 7 pA of beam current), we have not observed any electron-beam induced degradation~\cite{taylor2022hyperspectral}. While environmental degradation limited CL characterization of the pure FAPbI$_3$ and MAPbI$_3$ films (see \SM~Sec.~I for more data), the hybrid film exhibited a bright CL band near 800 nm consistent with near band-edge luminescence.

The correlated SE and CL images acquired as a function of laser exposure time are shown in Fig.~\ref{fig:CL}. We used non-negative matrix factorization (NMF) to decompose the hyperspectral CL images into representative spectral components. The calculated explained variance ratio (EVR) is greater than 90\% with two spectral components for all of the spectrum images, and it is greater than 99\% with two spectral components for all of the images acquired after laser exposure. The two spectral components for each image are shown in the second row of Fig.~\ref{fig:CL}, while the spatial distribution of CL corresponding to the two spectral components are presented in row 3 and 4. 

A few critical features are visible in Fig.~\ref{fig:CL}. First, the surface morphology exhibits substantial microstructuring after laser exposure, and the size of the microstructured area increases with increasing laser exposure time.  Second, CL NMF decomposition component 1 is consistent with excitonic luminescence, and component 2 is consistent with near-band-edge CL from an intermediate phase that results from partial degradation of the film~\cite{xiao2015mechanisms,taylor2022hyperspectral}. Critically, we did not observe any substantial CL band from PbI$_2$ near 520 nm to 540 nm, and we did not observe any statistically significant increase in the intensity of the intermediate phase CL as a function of laser exposure or position. However, while the band-edge luminescence is completely suppressed in the microstructured area, the band-edge CL intensity increases by up to a factor of 20 in the bright ring surrounding the laser spot (with a monotonic increase in CL intensity as a function of laser exposure time). Thus, we can conclude that (a) the ion migration induced by laser exposure does not introduce any near-surface PbI$_2$ or partially decomposed hybrid perovskite phases with any statistical significance, (b) the ion migration completely suppresses band-edge luminescence inside the bright ring, and (c) the ion migration substantially enhanced the luminescence intensity within the bright ring in a manner that is determined by the laser exposure time. We observed similar results using a continuous-wave laser source at 532 nm for light exposure (see \SM~Sec.~I for further details). Interestingly, we also observed a partial recovery of the excitonic CL emission near 800 nm in an environmentally degraded FAPbI$_3$ sample after light exposure (see Fig.~S10 in the \SM), suggesting a beneficial impact of light-induced ion migration in degraded OIHP samples.

\subsection{Summary}
In summary, we utilized an \textit{in situ} laser source incorporated into an SEM to directly map lateral and vertical photo-induced ion migration pathways in OIHPs with SEI, EDS and CL. We observed long-ranged iodine migration far beyond the illuminated region, a surprising vertical migration of lead ions and cross-boundary migration of silicon and tin ions from the substrate. CL analysis indicated significantly enhanced band-edge emission in a bright ring surrounding the laser spot with complete suppression of band-edge emission within the ring.  No increase in luminescence due to partially degraded perovskite phases or to PbI$_2$ band-edge emission was observed. These results highlight the ubiquity of photo-induced ion migration processes in OIHPs and their potential impact on the material quality and device performance. While this understanding of photo-induced ion migration is certainly critical to the development of mitigation strategies to suppress unwanted ion migration, the observation of 20-fold enhancement in band-edge luminescence also highlights an opportunity to improve perovskite device performance through controlled ion migration. Our work also demonstrated that multi-modal optoelectronic spectroscopies that combine laser and converged electron-beam excitations~\textit{in situ} with variable PE energies and interaction volumes can be a valuable tool to study microscopic ion transport in emerging materials. 

\section{Methods}
\subsection{Sample Fabrication}
This section presents the method of fabrication for the samples considered in this study. The FTO-coated glass substrates were washed with the detergent aqueous solution, deionized water, and 2-propanol in the ultrasonic bath for 15 mins sequentially. Then the FTO substrate was treated with UV-ozone for 20 mins just before the perovskite precursor deposition. To prepare the 1.5M MAPbI$_3$ precursor solution, equal molar ratios of methylammonium iodide (MAI, GreatCell Solar, 99.99\%) and PbI$_2$ (TCI, 99.99\%) were dissolved in dimethylformamide (DMF, Sigma-Aldrich, 99.8\%) and dimethyl sulfoxide (DMSO, Sigma-Aldrich, 99.8\%) with a volume ratio of 4:1. To prepare the 1.4M FAPbI$_3$ precursor solution, equal molar ratios of formanidium iodide (FAI, GreatCell Solar, 99.99\%), and PbI$_2$ were dissolved in DMF and DMSO with a volume ratio of 8:1. The perovskite precursor solution was spin-coated onto the FTO substrates at 4000 rpm for 30 s, and 0.6 ml of diethyl ether was dripped onto the substrate at the time of 15 s to the end. The as-spun films were annealed at 100 $^{\circ}$C for 10 mins and 150 $^{\circ}$C for 20 mins for MAPbI$_3$ and FAPbI$_3$, respectively. 

\subsection{SEM and EDS Measurement}

Secondary electron imaging and EDS measurements reported in this work was conducted inside a ThermoFisher Quanta 650 FEG SEM hosted at the University of California Santa Barbara. The secondary electrons were detected using a standard Everhart-Thornley detector, while the EDS spectra were collected using a ThermoFisher UltraDry 30 EDS detector and interpreted using the ThermoFisher Pathfinder software. A high vacuum of $1 \times 10^{-6}$ torr was maintained inside the SEM chamber for all measurements. The laser beam was generated from a Yb-doped-fiber laser (Clark-MXR IMPULSE) with a fundamental wavelength of 1030 nm, an average pulse width of 150 fs and a repetition rate of 5 MHz. The fundamental wavelength was converted to 515 nm using a BBO crystal and used as the source for light exposure. The laser beam was fed into the SEM chamber through a transparent viewport on the chamber wall.

A fresh area on the sample was used for every light exposure test. After the sample was exposed to light with a controlled intensity for a given amount of time, a secondary electron image was taken with an accelerate voltage of 5 keV and a beam current of 300 pA. A low beam current was used to minimize electron-beam-induced damage to the sample. A dwell time of 1 $\mu$s at each pixel was used to form the image. Then EDS line scans across the center of the light-exposure area at every $\sim 7$ $\mu$m were taken with a beam current of 300 pA. 100 linescans (100 ms per scan) were taken and averaged to minimize electron-beam induced damage, while enhancing the signal-to-noise ratio. The change in the absolute count of X-ray photons emitted from the exposed area was calculated as compared to the pristine background area.  

\subsection{Cathodoluminescence}

Cathodoluminescence (CL) measurements were performed with a Delmic Sparc CL collection module in a ThermoFisher Quattro environmental SEM hosted in the Center for Nanophase Materials Sciences (CNMS) at the Oak Ridge National Laboratory. A parabolic mirror was used to collect CL generated under electron-beam excitation and to deliver the laser to the sample.  A retractable mirror was inserted to direct free-space coupled laser sources to the parabolic mirror and retracted to allow for in situ CL characterization using an Andor Kymera spectrograph. Either a variable power 532 nm Cobolt CW laser source or a 495 nm pulsed laser source generated from the second harmonic of a Mai Tai Ti:Sapphire laser with 100 fs pulse duration, 80 MHz repetition rate, and 1.25 mW average power were focused to a spot size of 5 microns by the parabolic mirror. After variable laser exposure times with the retractable mirror inserted, the mirror was retracted, and CL spectrum images were acquired with the electron-beam conditions described in the manuscript. Because the SEM and the adjacent optics table are independently floated, an Aligna active beam stabilization system was used to dynamically correct for any movement of the SEM relative to the optics table. 
Blind non-negative matrix factorization (NMF) performed with scikit-learn~\cite{pedregosa2011scikit} was used to track changes in the CL spectrum images as a function of laser exposure time.

\bibliography{references.bib}

\begin{acknowledgments}
The work conducted at University of California Santa Barbara (SEI and EDS) was based on research supported by US Department of Energy, Office of Basic Energy Sciences, under the award number DE-SC0019244 (for the development of the laser-coupled SEM) and by the US Army Research Office under the award number W911NF-19-1-0060 (for studying photo-induced physics in emerging materials). Cathodoluminescence microscopies were conducted as part of a user project at the Center for Nanophase Materials Sciences (CNMS), which is a US Department of Energy, Office of Science User Facility at Oak Ridge National Laboratory. The sample fabrication at National Renewable Energy Laboratory (NREL) was supported by the US Department of Energy under Contract No. DE-AC36-08GO28308 with Alliance for Sustainable Energy, Limited Liability Company (LLC), the Manager and Operator of NREL. The authors at NREL also acknowledge the support on perovskite sample preparation from DE-FOA-0002064 and DE-EE0008790, funded by the US Department of Energy, Office of Energy Efficiency and Renewable Energy, Solar Energy Technologies Office. The views expressed in the article do not necessarily represent the views of the DOE or the U.S. Government.
\end{acknowledgments}

\section*{Author Contributions}

B. Liao, K.Z. and B. Lawrie conceived and supervised the project. S.P., Q.J. and K.Z. prepared the samples. T.K. conducted the SEI and EDS measurements at UCSB. U.C. and R.G. contributed to the instrument development at UCSB. V.I. and B. Lawrie conducted SEI and CL measurements at ORNL. G.E. and B. Lawrie conducted the CL data analysis. T.K., B. Liao and B. Lawrie drafted the manuscript. All authors have commented and edited the manuscript.  

\section*{Competing Interests}

The authors declare no competing interests.


\end{document}


\title{Supporting Information: Mapping the Pathways of Photo-induced Ion Migration in Organic-inorganic Hybrid Halide Perovskites}
\author{Taeyong Kim}
\affiliation{Department of Mechanical Engineering, University of California, Santa Barbara, CA 93106, USA}
\affiliation{Department of Mechanical Engineering, Seoul National University, South Korea}

\author{Soyeon Park}
\affiliation{National Renewable Energy Laboratory, Golden, CO 80401, USA}

\author{Vasudevan Iyer}
\affiliation{Center for Nanophase Materials Sciences, Oak Ridge National Laboratory, Oak Ridge, TN 37830, USA}

\author{Qi Jiang}
\affiliation{National Renewable Energy Laboratory, Golden, CO 80401, USA}

\author{Usama Choudhry}
\affiliation{Department of Mechanical Engineering, University of California, Santa Barbara, CA 93106, USA}

\author{Gage Eichman}
\affiliation{Center for Nanophase Materials Sciences, Oak Ridge National Laboratory, Oak Ridge, TN 37830, USA}

\author{Ryan Gnabasik}
\affiliation{Department of Mechanical Engineering, University of California, Santa Barbara, CA 93106, USA}

\author{Benjamin Lawrie}
\email{lawriebj@ornl.gov}
\affiliation{Center for Nanophase Materials Sciences, Oak Ridge National Laboratory, Oak Ridge, TN 37830, USA}
\affiliation{Materials Science and Technology Division, Oak Ridge National Laboratory, Oak Ridge, TN 37830, USA}

\author{Kai Zhu}
\email{kai.zhu@nrel.gov }
\affiliation{National Renewable Energy Laboratory, Golden, CO 80401, USA}

\author{Bolin Liao}
\email{bliao@ucsb.edu} \affiliation{Department of Mechanical Engineering, University of California, Santa Barbara, CA 93106, USA}


\date{\today}

{    \global\let\newpagegood\newpage
    \global\let\newpage\relax
\maketitle}
\clearpage

\section{Additional Data}

\subsection{Additional SE Images with Different Optical Power}

Light exposure tests with lower optical power compared to those reported in the main text were conducted and the results are shown here. Bright contrasts were still observed in both FAPbI$_3$ and MAPbI$_3$ with a reduced optical power. In the case of FAPbI$_3$, no bright-ring-shaped contrast was formed with the reduced optical power up to 3-hour exposure.

\begin{figure}[hbt!]
{
\includegraphics[width=\textwidth,height=\textheight/2,keepaspectratio]{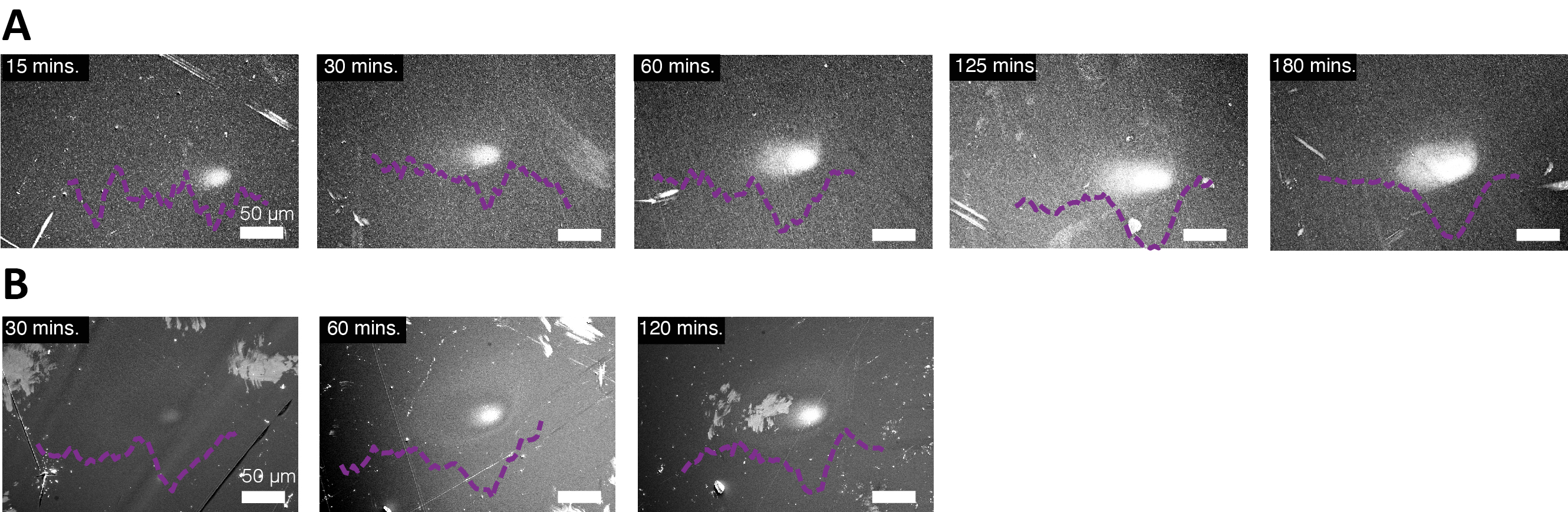}
\caption{\label{sfig:SEandEDS}
\textbf{Additional SE images and EDS horizontal line profiles of iodine at different optical fluences and exposure times.} (A) Measured data taken in FAPbI$_3$ at optical power of $\sim 2.5$ mW (optical fluence: $\sim 39.8$ $\mu$J~cm$^{-2}$), and (B) those in MAPbI$_3$ at optical power of $\sim 4.1$ mW (optical fluence: $\sim 66.0$ $\mu$J~cm$^{-2}$). Exposure time is labeled in the images.
}}
\end{figure}
\clearpage

\subsection{EDS spectra for pristine MAPbI$_3$}
Additional EDS spectra measured at primary electron (PE) energy of 30 keV for pristine MAPbI$_3$ is shown here.
\begin{figure}[hbt!]
{
\includegraphics[width=\textwidth,height=\textwidth/4,keepaspectratio]{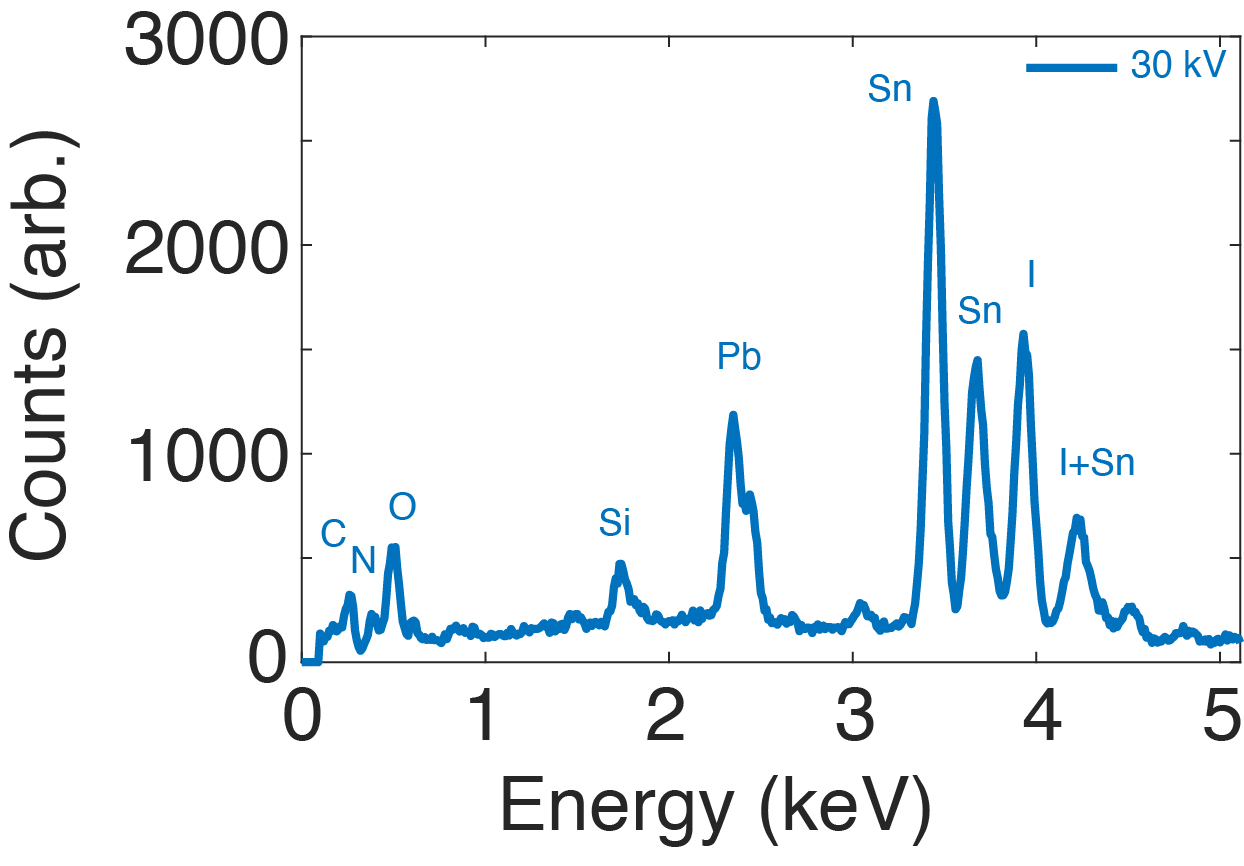}
\caption{\label{sfig:EDSspectrumMAPI}
\textbf{EDS spectrum of pristine MAPbI$_3$.} Shown is collected X-ray counts versus photon energy at PE energy of 30 keV in pristine MAPbI$_3$ without light exposure. Detected elements are annotated near each characteristic peak in the figure.
}}
\end{figure}

\clearpage

\subsection{EDS spectra at other PE energies for pristine MAPbI$_3$}
\begin{figure}[hbt!]
{
\includegraphics[width=0.8\textwidth,height=\textheight/2,keepaspectratio]{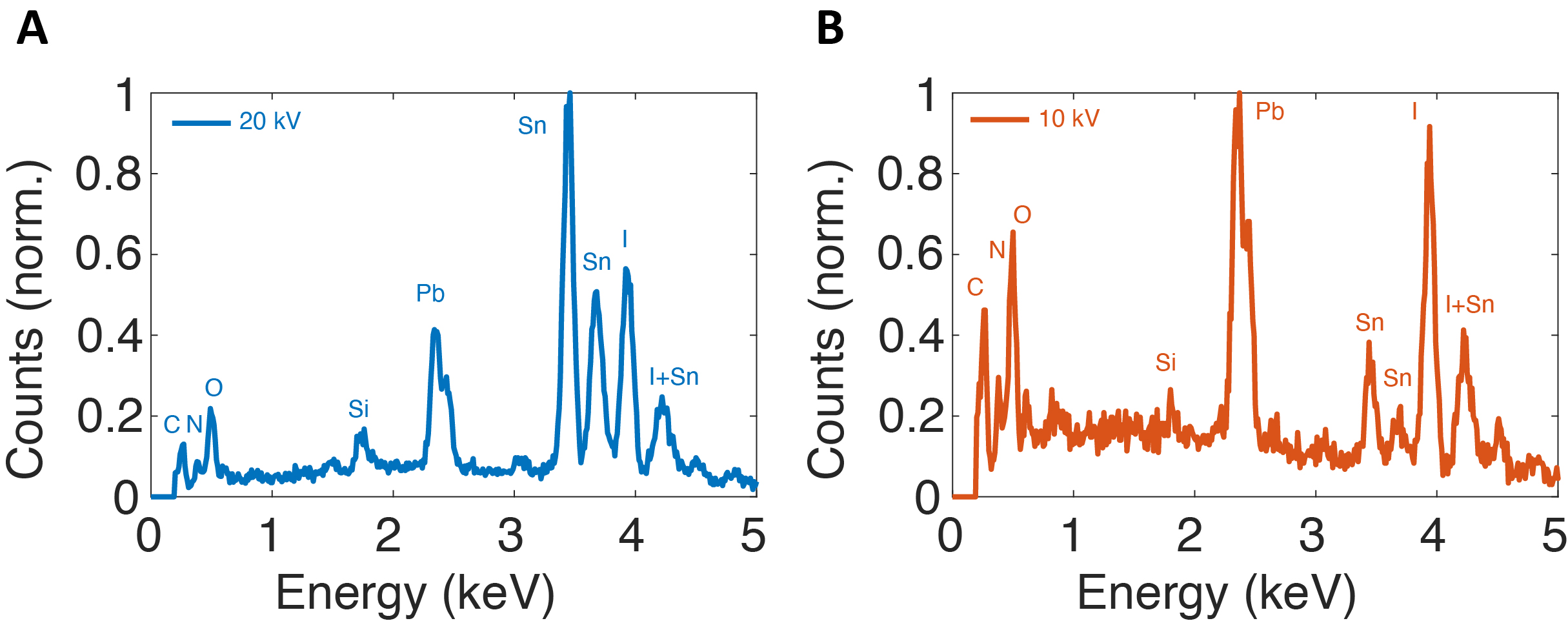}
\caption{\label{sfig:EDS20and10keV}
\textbf{Additional EDS emission spectrum at other PE energies for pristine MAPbI$_3$.} Measured EDS spectra at PE energies of (A) at 20 keV, and (B) at 10 keV. Detected elements are annotated near each characteristic peak in each figure.
}}
\end{figure}

In this section, additional EDS spectra measured at other PE energies are presented. The measured datasets are given in Fig.~\ref{sfig:EDS20and10keV}. The magnitude of the characteristic peaks due to the elements in the substrate (tin and silicon) decreases as the PE energy decreases, indicating that the corresponding EDS spectra depends on the penetration of the PEs through the depth of the sample.

\clearpage

\subsection{Additional EDS linescan results for FAPbI$_3$ and MAPbI$_3$}
\begin{figure}[hbt!]
{
\includegraphics[width=0.8\textwidth,height=\textheight/2,keepaspectratio]{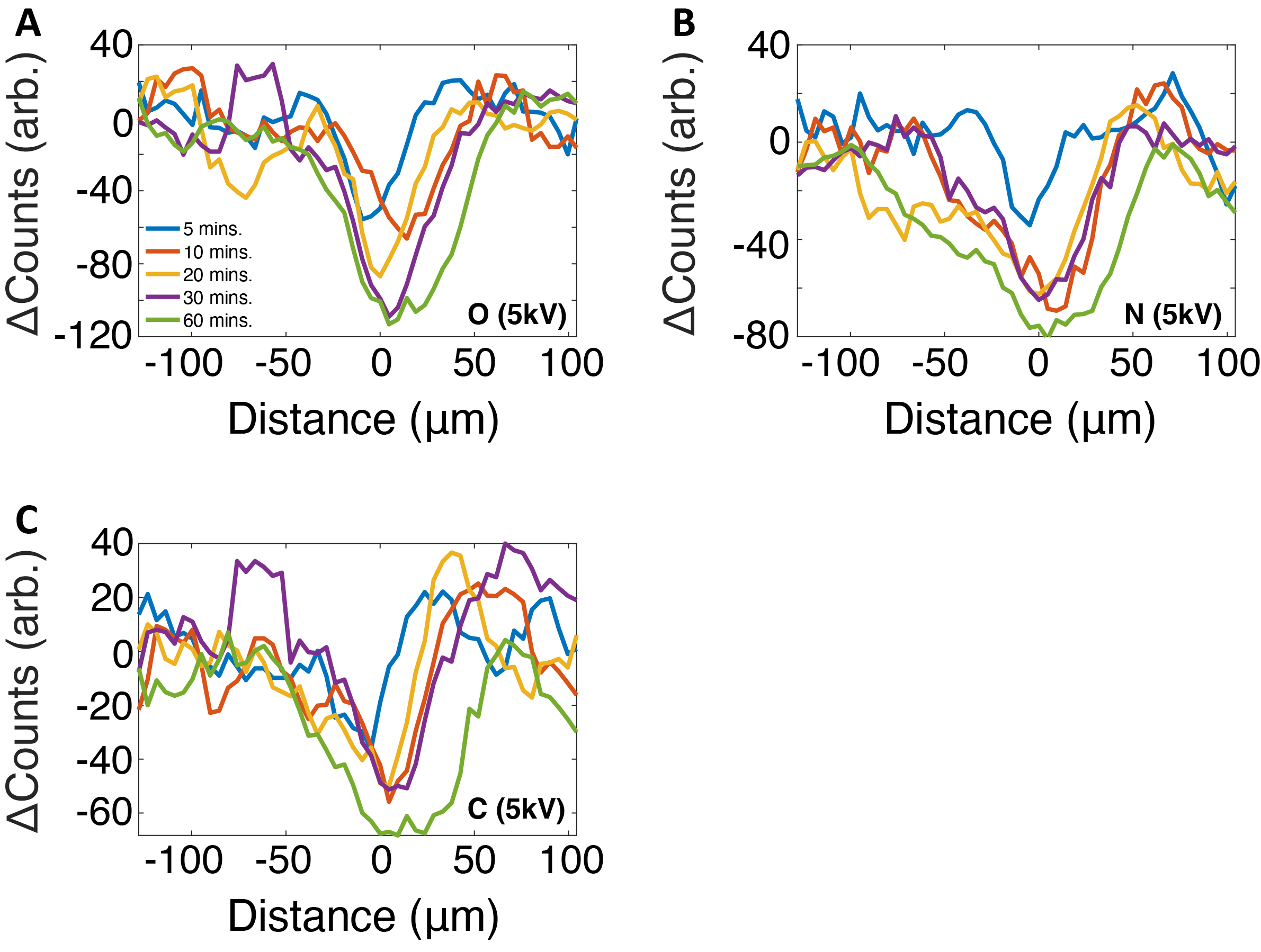}
\caption{\label{sfig:addEDS_7mW_FAPI}
\textbf{Additional EDS linescan results in FAPbI$_3$ at optical power of 6.7 mW, corresponds to an optical fluence of $\sim 106.6$ $\mu$J~cm$^{-2}$.} 
Measured X-ray counts versus distance from the center of the optical beam for (A) O (5-keV PE energy), (B) N (5-keV  PE energy), and (C) C (5-keV of PE energy). Legends indicate the optical exposure time. A reduction of these elements suggest a reduced concentration of the organic cations near the surface in the illuminated area, indicating photo-induced decomposition of the organic cations.
}}
\end{figure}

\begin{figure}[hbt!]
{
\includegraphics[width=0.8\textwidth,height=.8\textheight,keepaspectratio]{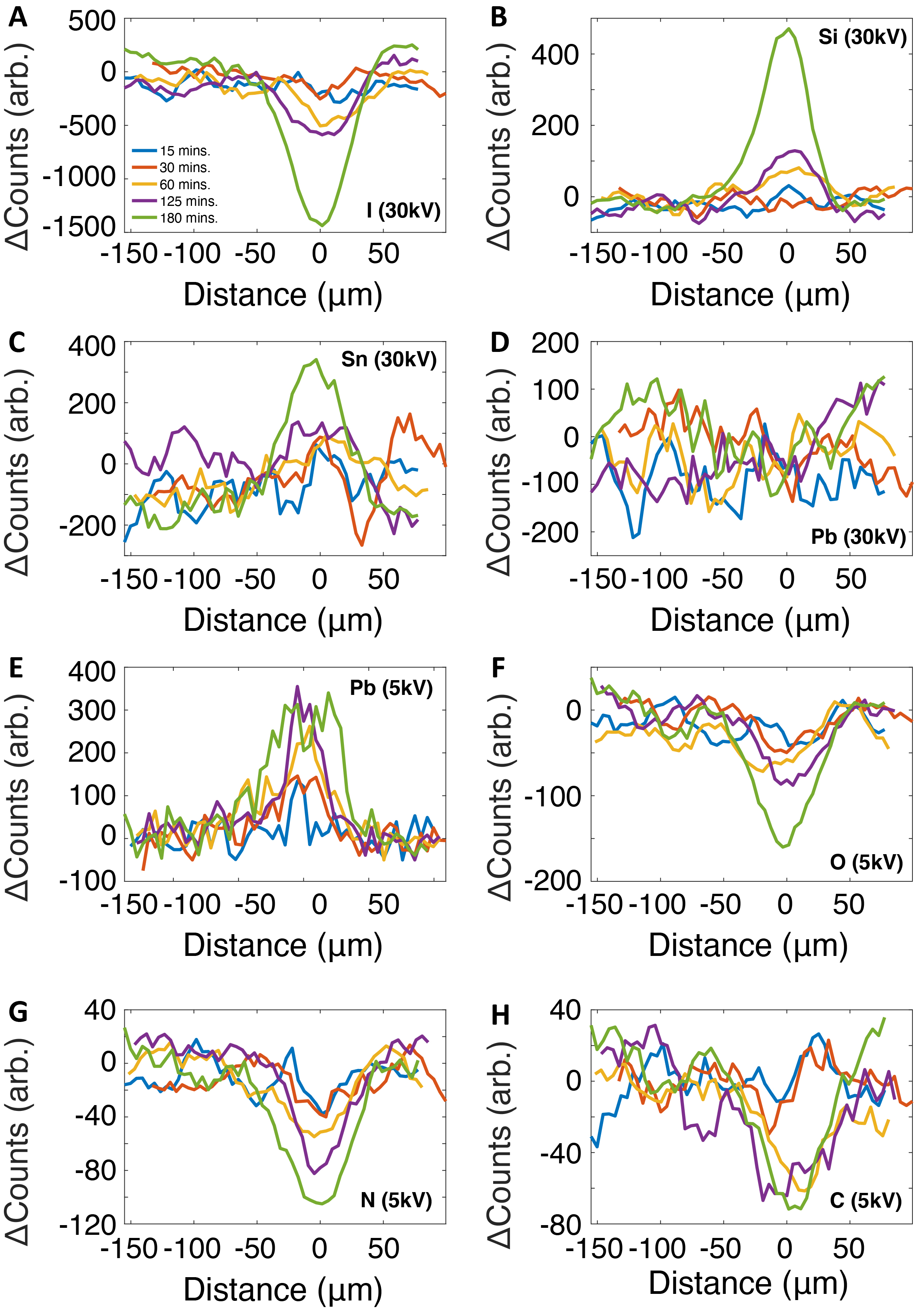}
\caption{\label{sfig:addEDS_3mW_FAPI}
\textbf{Additional EDS linescan results in FAPbI$_3$ at optical power of 2.5 mW, corresponding to an optical fluence of $\sim 39.8$ $\mu$J~cm$^{-2}$.} 
Measured X-ray counts versus distance from the center of the optical beam for (A) I (30-keV PE energy), (B) Si (30-keV PE energy), and (C) Sn (30-keV PE energy), (D) Pb (30-keV PE energy), (E) Pb (5-keV PE energy), (F) O (5-keV PE energy), (G) N (5-keV PE energy), and (H) C (5-keV PE energy). Legends indicate optical exposure time. The qualitative features of the ion distributions are similar to the case with a higher optical fluence that is discussed in the main text. 
}}
\end{figure}

\begin{figure}[hbt!]
{
\includegraphics[width=0.8\textwidth,height=.8\textheight,keepaspectratio]{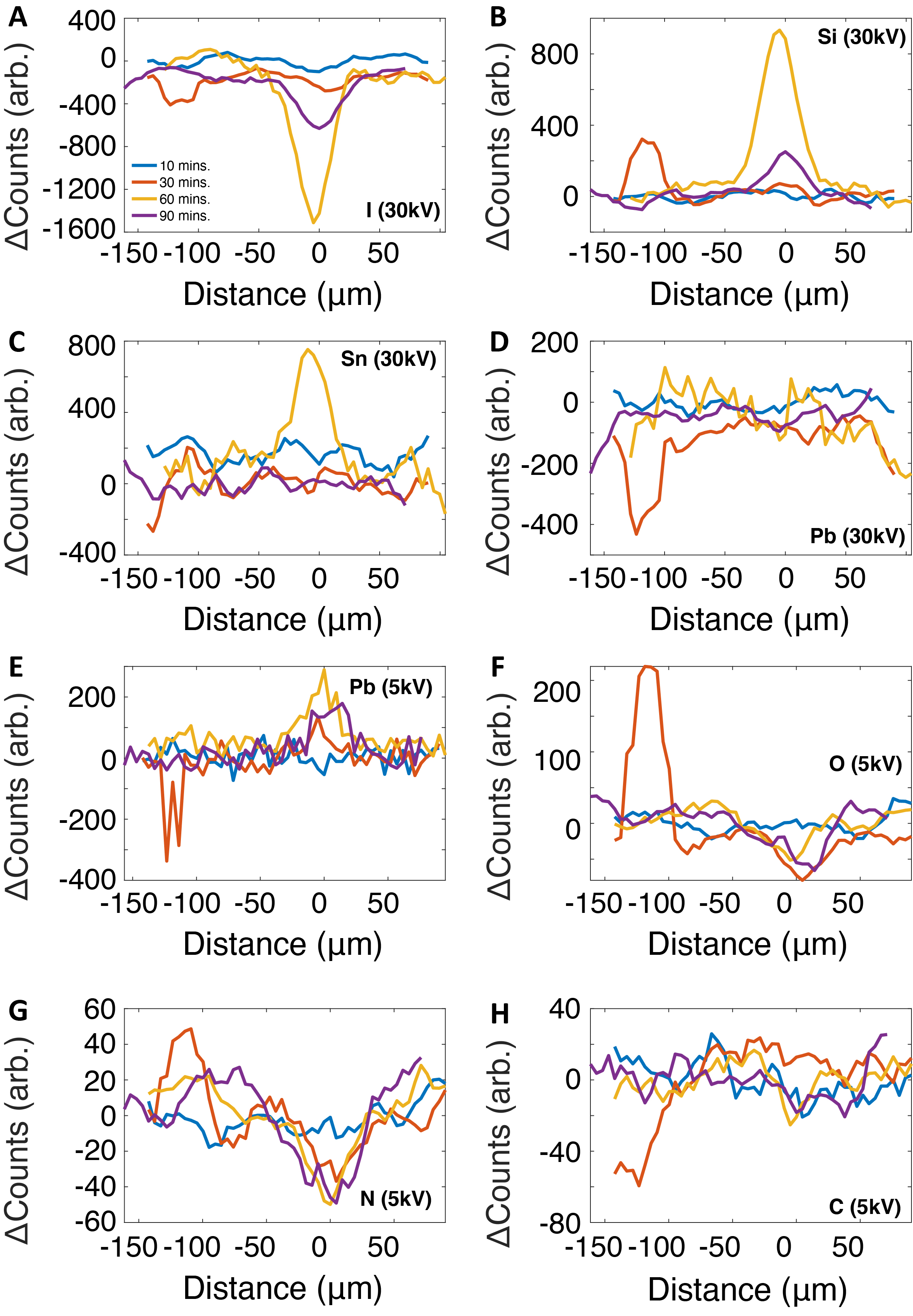}
\caption{\label{sfig:addEDS_7mW_MAPI}
\textbf{Additional EDS linescan results in MAPbI$_3$ at optical power of 6.7 mW, corresponding to an optical fluence of $\sim 106.6$ $\mu$J~cm$^{-2}$.} 
Measured X-ray counts versus distance from the center of the optical beam for (A) I (30-keV PE energy), (B) Si (30-keV PE energy), and (C) Sn (30-keV PE energy), (D) Pb (30-keV PE energy), (E) Pb (5-keV PE energy), (F) O (5-keV PE energy), (G) N (5-keV PE energy), and (H) C (5-keV PE energy). Legends indicate optical exposure time. The qualitative features of the ion distribution after light exposure in MAPbI$_3$ are similar to those observed in FAPbI$_3$ as discussed in the main text.
}}
\end{figure}

As discussed in the main text, we performed the EDS linescan experiments in FAPbI$_3$ and MAPbI$_3$ at various optical powers, exposure times, and PE energies. The corresponding additional data is presented in Figs.~\ref{sfig:addEDS_7mW_FAPI},~\ref{sfig:addEDS_3mW_FAPI},~\ref{sfig:addEDS_7mW_MAPI}. We note that an artifact near horizontal location of $\sim 100$ $\mu$m can be seen in Fig.~\ref{sfig:addEDS_7mW_MAPI}, which is due to a morphological defect on the scanned area in MAPbI$_3$.

\begin{figure}[hbt!]
{
\includegraphics[width=0.8\textwidth,height=.8\textheight,keepaspectratio]{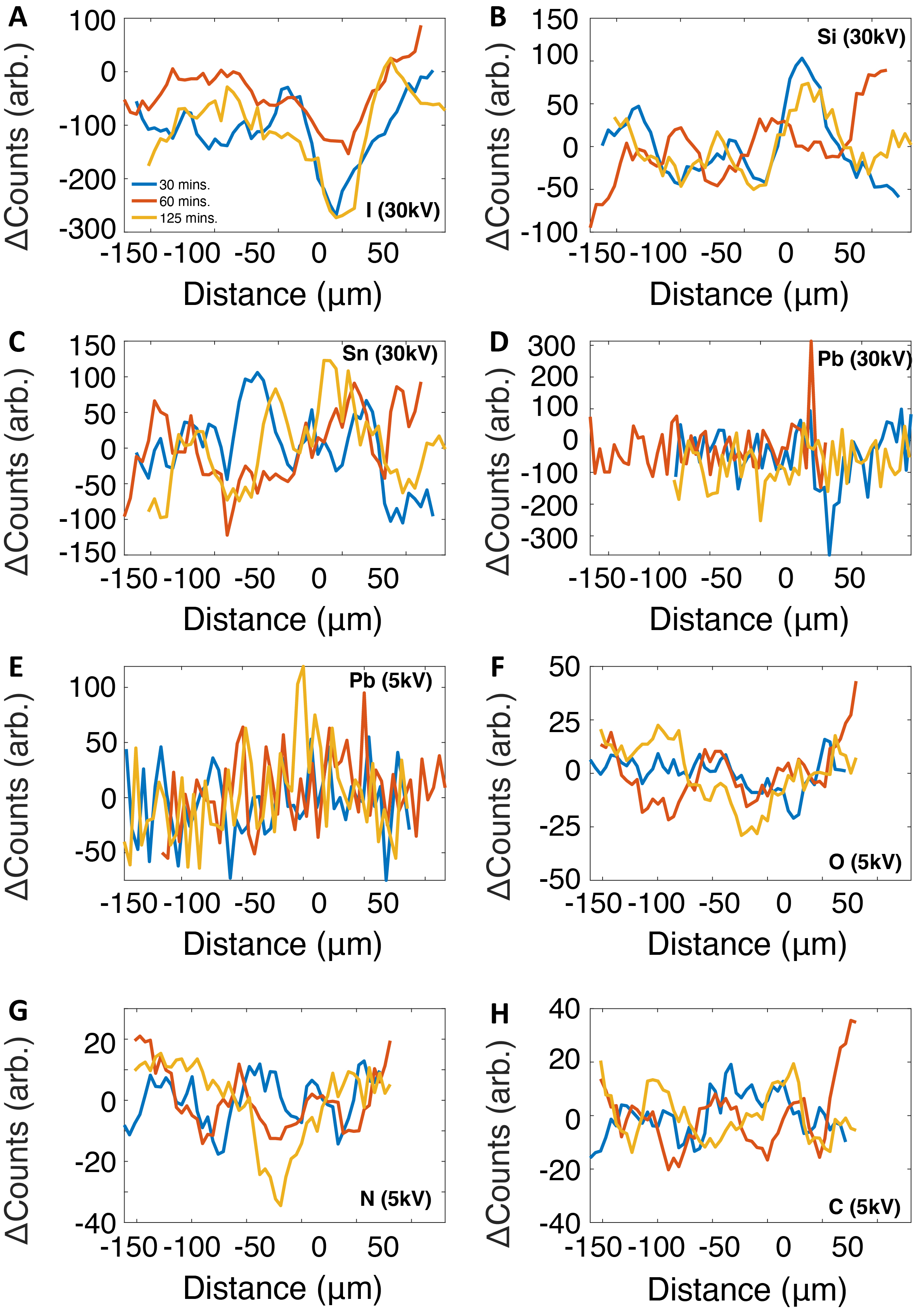}
\caption{\label{sfig:addEDS_4mW_MAPI}
\textbf{Additional EDS linescan results in MAPbI$_3$ at optical power of 4.1 mW, corresponding to an optical fluence of $\sim 66.0$ $\mu$J~cm$^{-2}$.} 
Measured X-ray counts versus distance from the center of the optical beam for (A) I (30-keV PE energy), (B) Si (30-keV PE energy), and (C) Sn (30-keV PE energy), (D) Pb (30-keV PE energy), (E) Pb (5-keV PE energy), (F) O (5-keV PE energy), (G) N (5-keV PE energy), and (H) C (5-keV PE energy). Legends indicate optical exposure time.
}}
\end{figure}

\clearpage

\subsection{EDS horizontal line profile of silicon at 5-keV under various optical exposure time.}
\begin{figure}[hbt!]
{
\includegraphics[width=0.8\textwidth,height=\textheight/2,keepaspectratio]{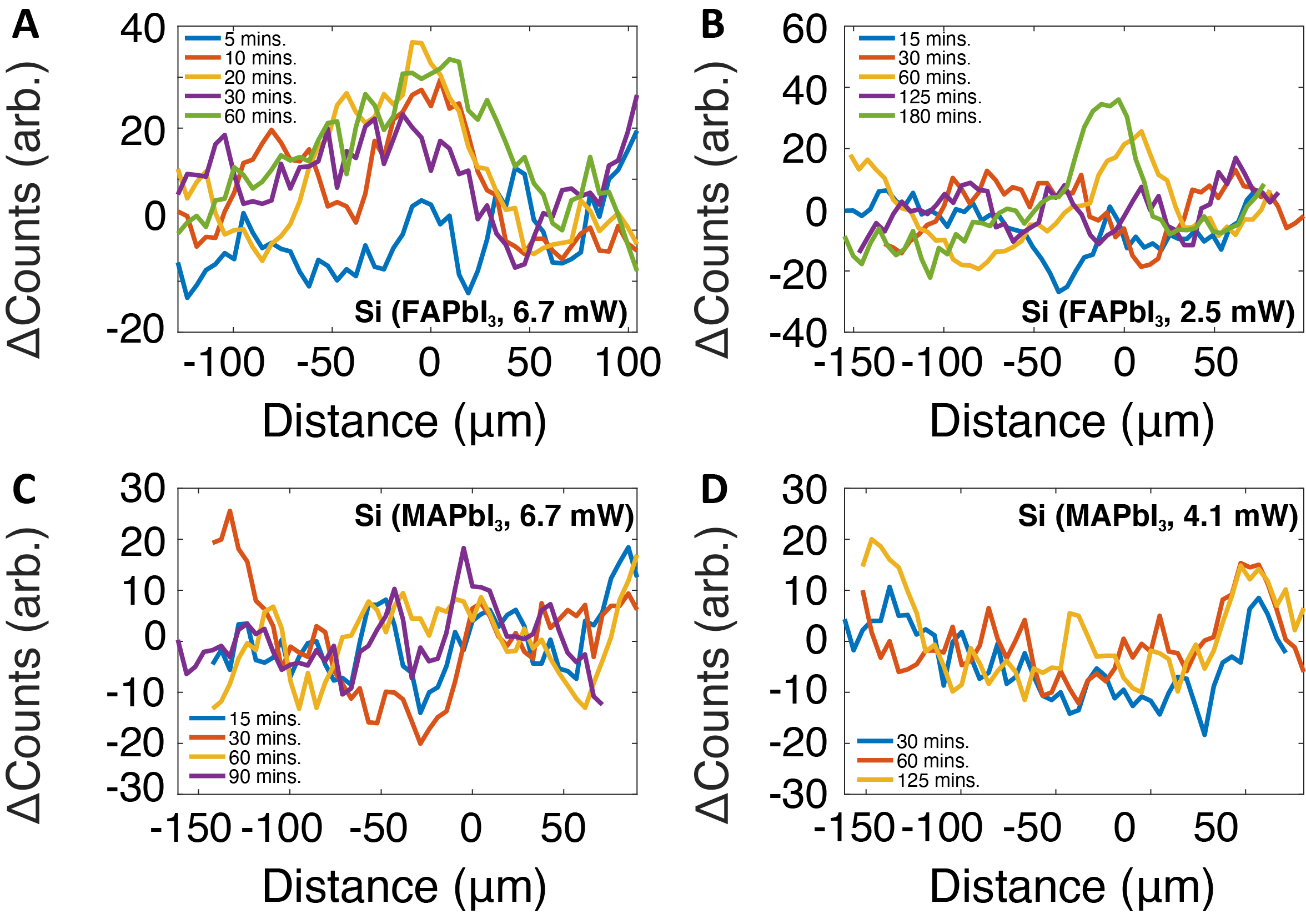}
\caption{\label{sfig:EDSSi}
\textbf{EDS linescan results for silicon at 5-kV in FAPbI$_3$ and MAPbI$_3$} Measured EDS linescan for silicon at optical power of (A) 6.7 mW (optical fluence: $\sim 106.6$ $\mu$J~cm$^{-2}$), (B) 2.5 mW (optical fluence: $\sim 39.8$ $\mu$J~cm$^{-2}$) in FAPbI$_3$. Also shown are those at optical power of (C) 6.7 mW (optical fluence: $\sim 106.6$ $\mu$J~cm$^{-2}$), and (D) 4.1 mW ($\sim 66.0$ $\mu$J~cm$^{-2}$) in MAPbI$_3$. Legends indicate the optical exposure time.
}}
\end{figure}

Additional EDS horizontal line profile for silicon at 5-keV PE energy under various optical exposure times and optical powers is shown in Fig.~\ref{sfig:EDSSi}. At 5-keV PE energy, despite the presence of detected small trace of silicon, their distribution did not change appreciably due to the light exposure, suggesting the migration of silicon ions mainly occurred deep inside the bulk sample.

\clearpage

\subsection{Cathodoluminescence Data after CW Laser Exposure}
\begin{figure}[hbt!]
{
\includegraphics[width=0.7\textwidth]{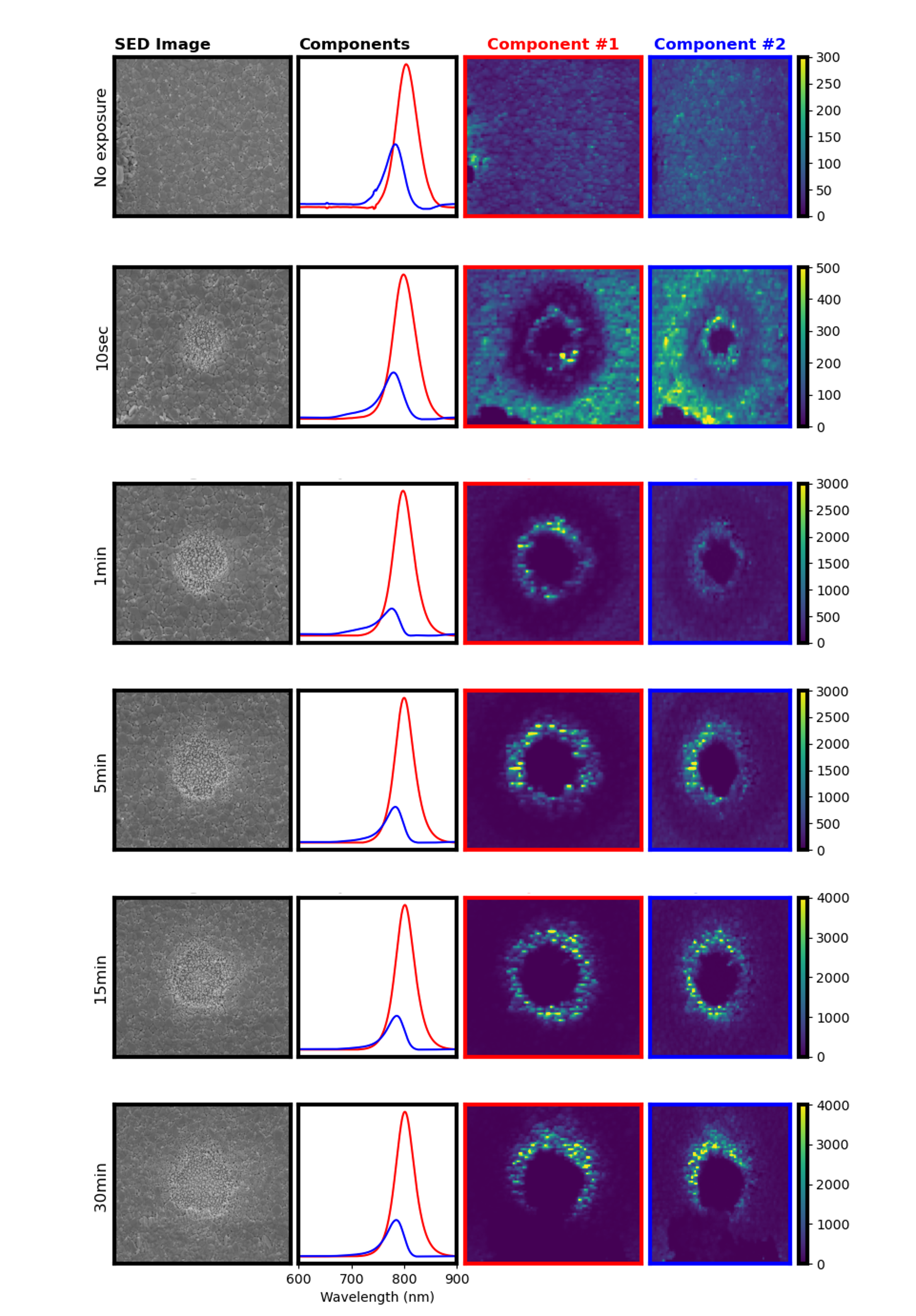}
\caption{\label{sfig:cl}
\textbf{Cathodoluminescence analysis of 97\% FAPbI$_3$/3\% MAPbBr$_3$ after \textit{in situ} exposure with a 532 nm, CW, 3 mW CW laser source.} Concurrent SE images (first column) and CL NMF decompositions for a pristine sample and for samples exposed to the laser for 10 s, 1 minute, 5 minutes, 15 minutes, and 30 minutes. The two spectral components of each NMF decomposition are illustrated in the second column, and the intensity maps for each of those components are illustrated in the third and fourth columns. The color bars highlight the increase in band-edge CL intensity as a function of laser exposure time. All images have a 22.2 $\mu$m horizontal field width.
}}
\end{figure}

Additional NMF decompositions of CL spectrum images acquired for a hybrid 97\% FAPbI$_3$/3\% MAPbBr$_3$ perovskite film exposed to an \textit{in situ} 532 nm, CW, 3 mW laser focused to a 5 $\mu$m spot are shown in Fig.~\ref{sfig:cl}. The results are highly consistent with the CL spectrum images acquired after pulsed laser excitation and illustrated in Fig. 5 of the main text. In particular, the perovskite band-edge luminescence is completely suppressed under laser excitation, but a bright ring with up to an order of magnitude enhancement of band-edge luminescence is observed surrounding the exposed area. Also, no PbI$_2$  band-edge luminescence is observed before or after laser exposure, and the laser exposure doesn't change the relative abundance of the intermediate degraded perovskite phase.

\clearpage

\subsection{Cathodoluminescence Data after CW Laser Exposure of environmentally degraded FAPbI$_3$}

\begin{figure}[hbt!]
{
\includegraphics[width=0.7\textwidth]{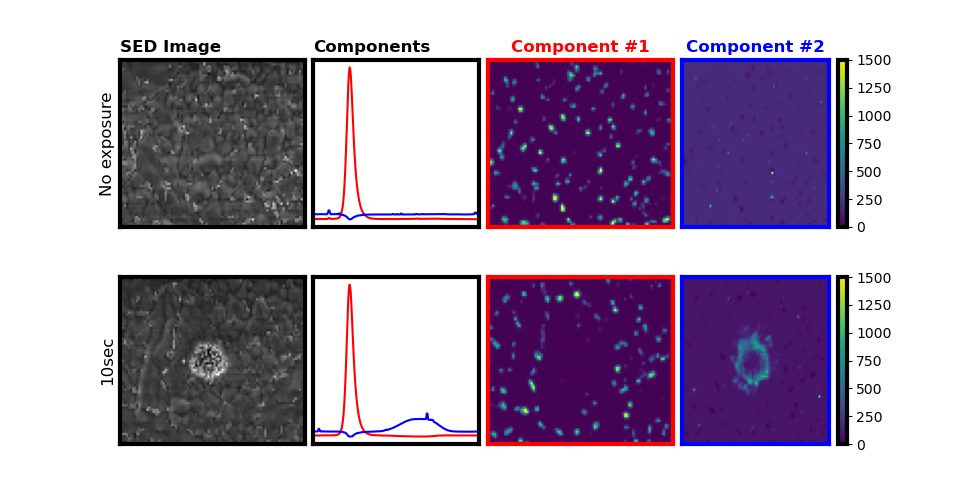}
\caption{\label{sfig:c2}
\textbf{Cathodoluminescence analysis of FAPbI$_3$ film after complete environmental degradation.} After extensive environmental degradation, the FAPbI$_3$ film exhibits only heterogeneous PbI$_2$ luminescence with no band edge luminescence visible in the NMF decomposition of a CL spectrum image acquired prior to laser irradiation (first row).  After 10 s \textit{in situ} exposure with a 532 nm, CW, 3 mW CW laser source, the PbI$_2$ luminescence near the laser spot is completely suppressed, and the perovskite luminescence is partially recovered, as seen in component two of the NMF reconstruction (second row).  Both datasets have a 14.6 $\mu$m horizontal field width.
}}
\end{figure}

After extended exposure to ambient conditions, FAPbI$_3$ and MAPbBr$_3$ films exhibit substantial degradation, with no visible band-edge luminescence remaining. The first row of Fig.~\ref{sfig:c2} illustrates a prototypical CL image of an environmentally degraded perovskite film (in this case, a FAPbI$_3$ film).  Bright PbI$_2$ luminescence is visible in localized regions (component 1), but no band edge perovskite luminescence is visible.  Intriguingly, as seen in the second row of Fig.~\ref{sfig:c2}, after 10s exposure to a CW, 532 nm, 3 mW laser source, the PbI$_2$ luminescence near the exposed area is completely suppressed, and a bright ring of FAPbI$_3$ luminescence around the exposed area emerges.  This result points to the potential impact of laser induced ion migration in perovskite thin films as a resource for remedying environmental degradation.

\section{Monte Carlo simulation of primary electron penetration through the specimen}
\begin{figure}[hbt!]
{
\includegraphics[width=0.75\textwidth,height=\textheight/2,keepaspectratio]{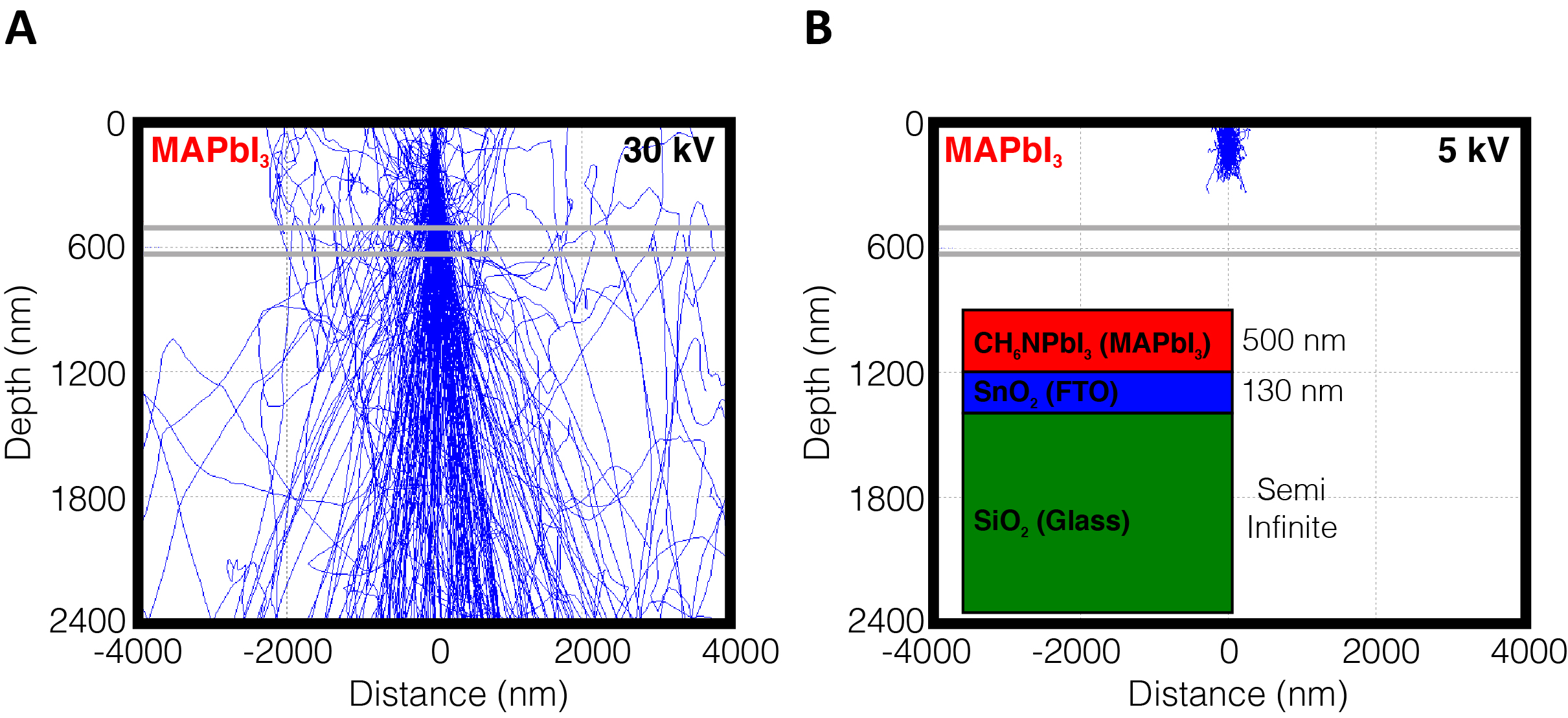}
\caption{\label{sfig:elpath}
\textbf{Monte Carlo Simulation of Electron Trajectory in MAPbI$_3$.} The simulated trajectories of primary electrons with energies of (A) 30 keV, and (B) 5 keV in a MAPbI$_3$ sample. While the primary electrons can penetrate the entire sample stack at 30 keV, those at 5 keV are mostly confined within 200 nm near the MAPbI$_3$ surface. The interfaces between MAPbI$_3$ and FTO, and between FTO and the glass substrate are labeled with gray lines. The inset in (B) shows the structure of the MAPbI$_3$ sample stack.
}}
\end{figure}

We simulated the trajectory of primary electrons across the sample considered in this study using the computer program CASINO (Monte Carlo Simulation of Electron Trajectory in Solids)~\cite{Hovington_CASINO}. As given in the inset of Fig.~\ref{sfig:elpath}B, we considered our sample geometry containing three layers (two thin layers and one thick layer). The corresponding thin layers are 500-nm MAPbI$_3$ or 800-nm FAPbI$_3$, and 130-nm SnO$_2$. For modeling purposes, we treated the SiO$_2$ substrate as a semi-infinite, despite its actual thickness of $\sim 3$~mm. Then, we assumed that the electron beam (total number of electrons: 10000, electron beam radius: 10 nm) was incident on the specimen with initial energy of 30 keV and 5 keV, depending on the primary electron acceleration voltages in our SEM settings, and that they underwent successive collision events inside the specimen. Figure~\ref{sfig:elpath}A shows our simulation results for MAPbI$_3$ layer in which many of the primary electrons penetrate through the entire MAPbI$_3$ layer and into the substrate at 30 keV. In contrast, the primary electrons with initial kinetic energies of 5 keV were mostly confined within $\lesssim 200$ nm near the surface. The simulation results of FAPbI$_3$ is shown in Fig.~1C and D in the main text.

\clearpage

\section{Discussion of fitting procedures for determining diffusivity}
\begin{figure}[hbt!]
{
\includegraphics[width=0.75\textwidth,height=\textheight/2,keepaspectratio]{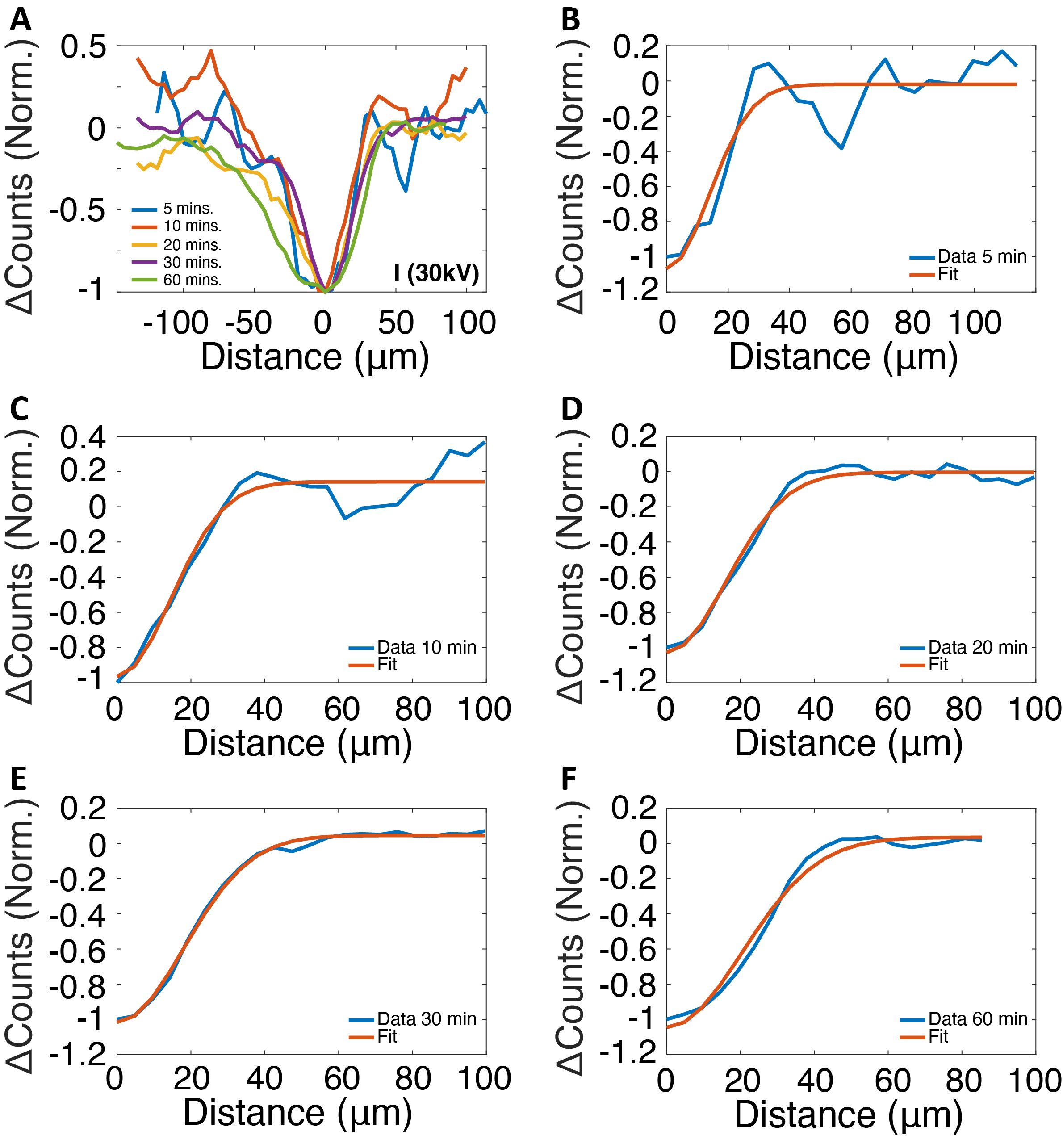}
\caption{\label{sfig:fitting_right}
\textbf{Analysis of horizontal EDS linescan profile for iodine using the right-half profiles (optical fluence: $\sim 106.6$ $\mu$J~cm$^{-2}$).} (A) The normalized difference counts versus distance at various optical exposure times ranging from 5 minutes to 60 minutes. The profile was divided into two at the horizontal origin, in which right fractional profiles under optical exposure for (B) 5 minutes, (C) 10 minutes, (D) 20 minutes, (E) 30 minutes, and (F) 60 minutes were analyzed (blue solid lines). The corresponding profile was fitted using one-dimensional  Gaussian models (red solid lines), in which we found a reasonable agreement for all the data considered here.
}}
\end{figure}

\begin{figure}[hbt!]
{
\includegraphics[width=0.75\textwidth,height=\textheight/2,keepaspectratio]{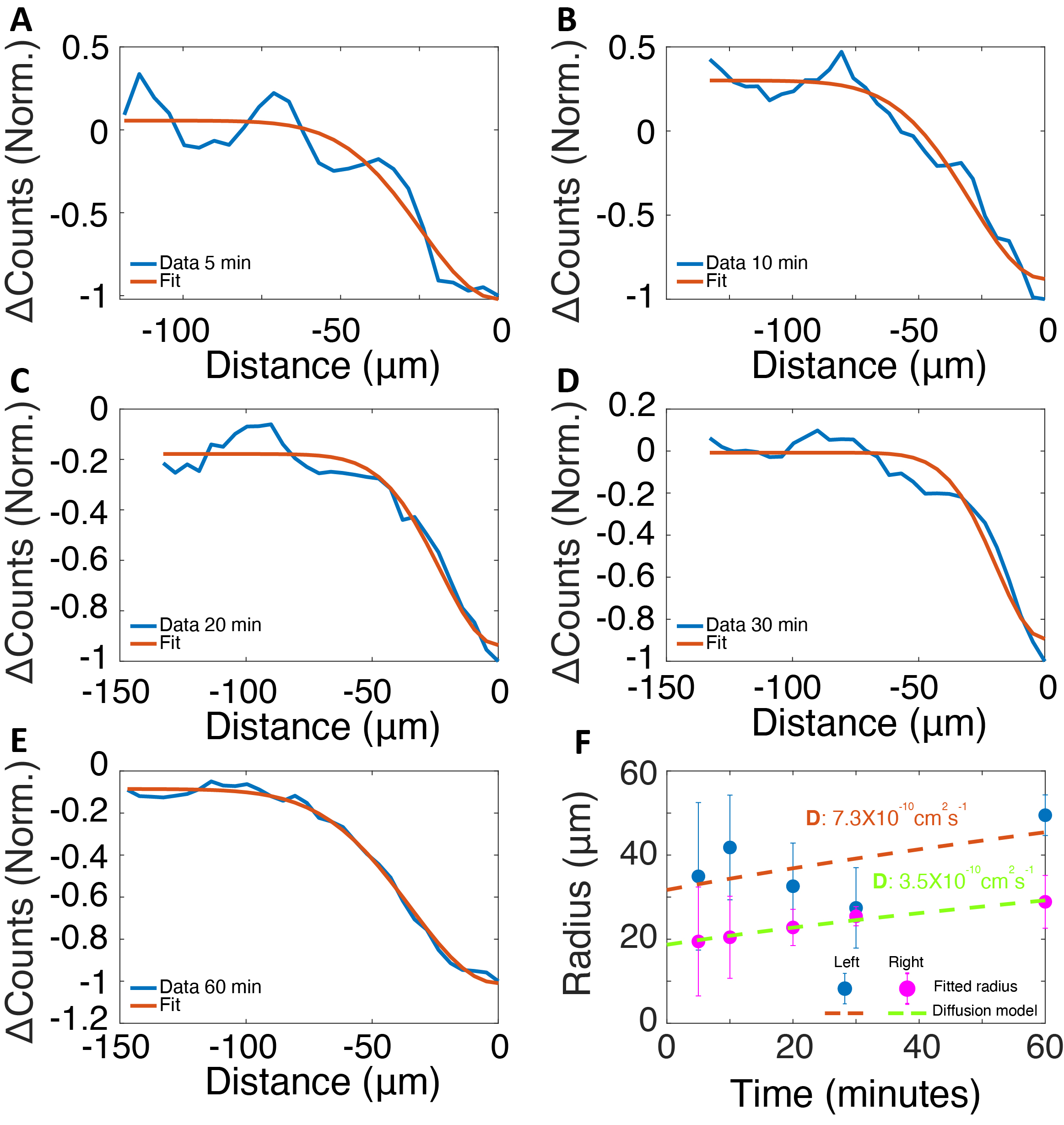}
\caption{\label{sfig:fitting_left}
\textbf{Analysis of horizontal EDS linescan profile for iodine using left-half profiles (optical fluence: $\sim 106.6$ $\mu$J~cm$^{-2}$).} The analyzed left fractional profiles under optical exposure for (A) 5 minutes, (B) 10 minutes, (C) 20 minutes, (D) 30 minutes, and (E) 60 minutes were analyzed (blue solid lines), along with one-dimensional  Gaussian model fits (red solid lines). The discrepancies between the data and Gaussian fit can be seen due to asymmetry of the profiles and background noise. (F) The Gaussian-fitted radii versus the optical exposure time determined using left and right profiles are shown. The error bars indicate numerically determined 95\% confidence intervals. The dashed lines indicate best fits determined using the one-dimensional diffusion model, from which the diffusivities are annotated near each fit.}}
\end{figure}

This section presents our procedures for determining the diffusivity of iodine ions. First, the profiles of the horizontal linescans were normalized, as shown in Fig.~\ref{sfig:fitting_right}A. Next, due to emerging asymmetry of the profiles and background noise, the normalized profiles were divided into right (Fig.~\ref{sfig:fitting_right}B-F) and left (Fig.~\ref{sfig:fitting_left}A-F) sections, which were separately analyzed using a one-dimensional Gaussian model. Although we carefully attempted to fit the left profiles in Fig.~\ref{sfig:fitting_left}A-F, we found that the data and fit exhibit slight discrepancies particularly near the origin, leading to more scattered resulting fitted radii, thus complicating our analysis, as shown in Fig.~\ref{sfig:fitting_left}F. In contrast, we found an improved agreement between the right profiles and the Gaussian fits, as shown in Fig.~\ref{sfig:fitting_right}B-F. As a consequence, the trend for the radii versus exposure time was less scattered compared to those for the left profiles (see Fig.~\ref{sfig:fitting_left}F). Then, once the fitted radii were determined at each exposure time, we additionally fitted the diffusivity assuming one-dimensional (horizontal) diffusion~\cite{Bolin_Natnano}. The results of fitting using the one-dimensional diffusion model are presented as dashed lines in Fig.~\ref{sfig:fitting_left}F. We here note that despite more scattered Gaussian-fitted radii emerging for the left profiles, the fitted values of the diffusion coefficients from left and right profiles were reasonably close to each other. We also note that the calculation here is meant to be an order-of-magnitude estimation, since the real ion migration process is more complicated than the simple one-dimensional diffusion model used here. 

\clearpage

\section{Estimation of X-ray self-absorption depth}
This section presents our estimation of the X-ray self-absorption depth. First, we consider the elemental specific energy bands of the X-rays emitted from the specimen, which is presented in Tab.~\ref{tab:xrayband}. Then, given the particular energy band, the total mass absorption coefficient by a certain element ($\mu/\rho$) was obtained using an established database \cite{henke1993x}, where $\mu$ is the linear absorption coefficient with a dimension of $1/$length (the inverse of the absorption depth) and $\rho$ is the elemental density. For a pure elemental material, the total atomic absorption cross section $\sigma_a$ can be calculated from the mass absorption coefficient by:
\begin{equation}
    \sigma_a = \frac{A}{N_A} \frac{\mu}{\rho},
    \label{eqn:xraycs}
\end{equation}
where $A$ is the atomic weight and $N_A$ is the Avogadro's constant. For a compound material, the mass absorption coefficient can be obtained from the sum of the absorption cross sections of the constituent atoms by~\cite{thompson2001x}:
\begin{equation}
    \left( \frac{\mu}{\rho} \right)_\mathrm{compound}=\frac{N_A}{A_\mathrm{tot}}\sum_i x_i \sigma_{ai},
    \label{eqn:xraytotal}
\end{equation}
where the molecular weight of a compound containing $x_i$ atoms of type $i$ is $A_\mathrm{tot}=\sum_i x_i A_i$. This equation does not include interactions among the constituent atoms, but is generally applicable for X-ray photon energies above 30 eV and sufficiently far from absorption edges~\cite{thompson2001x}. Given their large atomic masses, lead and iodine are the major contributors to the total absorption cross section in both MAPbI$_3$ and FAPbI$_3$. Combining Eqns.~\ref{eqn:xraycs} and \ref{eqn:xraytotal} and data from the database, the absorption lengths of the characteristic X-ray energies in MAPbI$_3$ were calculated and listed in Table~\ref{tab:xrayband}. The values in FAPbI$_3$ should be similar given the same lead and iodine composition. As shown in Table~\ref{tab:xrayband}, for lead, iodine and tin, the X-ray absorption depth far exceeds the sample thickness ($\sim 500$ nm for FAPbI$_3$; $\sim 800$ nm for MAPbI$_3$). Therefore, the probing depth of the EDS measurement of these elements was only determined by the penetration depth of the primary electrons. On the other hand, the X-ray absorption depths of carbon, nitrogen and oxygen were smaller than the sample thickness, and thus the EDS probe was only sensitive to their distributions near the sample surface. The situation of silicon is in the middle: even with 30-keV primary electrons, the EDS was only sensitive to silicon distribution within roughly 400 nm near the surface due to the fact that X-rays emitted deeper inside the sample cannot escape and be collected by the EDS detector.

\begin{table}[h]
 \caption{
Element specific X-ray emission bands detected in our experiments and their absorption length in MAPbI$_3$
 }
 \label{tab:xrayband}
 \begin{ruledtabular}
 \begin{tabular}{cccc}
Element & Photon energy (keV) & Characteristic principal line & Absorption length in MAPbI$_3$ (nm)\\ 
 \hline
Pb & 2.35 & M$_{\alpha1}$ & 854\\
I & 3.94 & L$_{\alpha1}$ & 2780\\
C & 0.277 & K$_\alpha$ & 148\\
N & 0.392 & K$_\alpha$ & 173\\
O & 0.525 & K$_\alpha$ & 264\\
Si & 1.74 & K$_\alpha$ & 424\\
Sn & 3.44 & L$_{\alpha1}$ &1970\\
\end{tabular}
 \end{ruledtabular}
\end{table}
\clearpage




\section{Estimation of laser-induced temperature rise}

Since we used a pulsed laser source in our experiment, we need to estimate both the transient temperature rise after each laser pulse is absorbed and the steady-state temperature rise induced by the periodic pulse train. The transient temperature rise can be estimated in the following way. The highest optical fluence used in our experiment was $\sim 100 \mu$J~cm$^{-2}$. The reflectance of MAPbI$_3$ and FAPbI$_3$ is about 25\% at 515 nm~\cite{shirayama2016optical,kato2017universal}. Given the optical absorption depth of roughly 100 nm in both MAPbI$_3$ and FAPbI$_3$ ~\cite{kim2020high,shirayama2016optical,kato2017universal} at 515 nm, the volumetric heat capacity of MAPbI$_3$ ($\sim 1.28 \times 10^{6}$ J~m$^{-3}$K$^{-1}$)~\cite{haeger2020thermal}, the transient temperature rise after each pulse is absorbed can be estimated to be:
\begin{equation*}
\Delta T_{\mathrm{transient}}=\frac{0.75 \times 100\,\mu \mathrm{J~cm}^{-2}}{100\,\mathrm{nm} \times 1.28 \times 10^6 \, \mathrm{J~m}^{-3}\mathrm{K}^{-1}}=5.8\, \mathrm{K}.    
\end{equation*}
    
\begin{figure}[hbt!]
{
\includegraphics[width=\textwidth,height=\textheight/2,keepaspectratio]{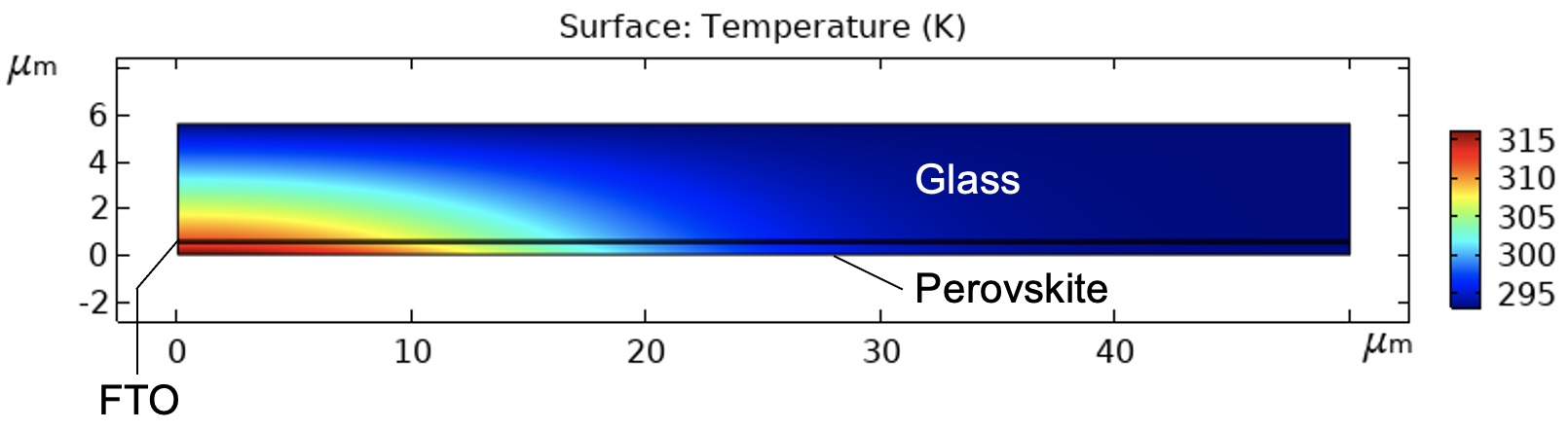}
\caption{\label{sfig:sstemp}
\textbf{Finite element simulation of the laser-induced steady-state temperature rise.} The simulated sample stack consisted of 500-nm MAPbI$_3$, 100-nm FTO and 5-$\mu$m glass. The environmental temperature was set at 293 K. The temperature distribution is coded in the color bar.}}
\end{figure}
    
We estimated the steady-state temperature rise using a finite element simulation (COMSOL Multiphysics). We used the thermal conductivity values for FTO: 4 W~m$^{-1}$K$^{-1}$~\cite{olson2018size}, and glass substrate: 1.2 W~m$^{-1}$K$^{-1}$. The steady-state temperature rise was obtained by simulating the heating effect of a continuous-wave source with the same average power (6.7 mW was the highest optical power used in our study). An optical beam $1/e^2$ radius of 25 $\mu$m and an optical absorption depth of 100 nm were used in the simulation. The simulated sample stack consisted of 500-nm MAPbI$_3$, 100-nm FTO and semi-infinite glass (5-$\mu$m thick in the simulation). The environmental temperature was set at 293 K. The simulated temperature distribution is shown in Fig.~\ref{sfig:sstemp}. From the simulation, the highest temperature at the center of the laser beam reached 316 K, which was 23 K higher than the initial temperature. Therefore, the overall laser-induced temperature rise (including both transient and steady state rises) is $\Delta T < 30$ K. This estimation suggests that the impact of laser heating to our observed ion migration is much smaller compared to a recent observation on the heat-induced migration effect, in which the temperature rise was on the order of $\sim 100$ degrees~\cite{lai2018intrinsic}.

\begin{figure}[hbt!]
{
\includegraphics[width=\textwidth,height=\textheight/2,keepaspectratio]{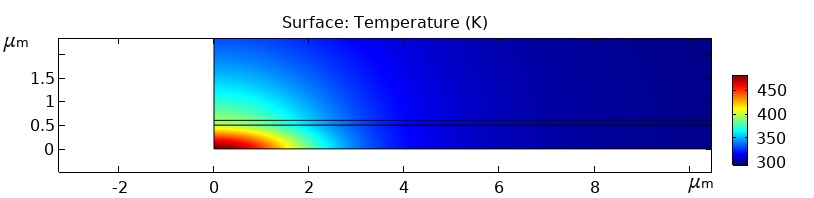}
\caption{\label{sfig:sstemp5um}
\textbf{Finite element simulation of the laser-induced steady-state temperature rise in the CL measurement.} The simulated sample stack consisted of 500-nm MAPbI$_3$, 100-nm FTO and 5-$\mu$m glass. The environmental temperature was set at 293 K. The optical beam diameter was 5 $\mu$m with an average power of 1.25 mW. The temperature distribution is coded in the color bar.}}
\end{figure}

For the CL measurement, due to the smaller beam size (5 $\mu$m diameter), the steady-state temperature rise was higher despite a similar optical fluence (average optical power was 1.25 mW and optical fluence was 80   $\mu$J~cm$^{-2}$. The simulated temperature rise was shown in Fig.~\ref{sfig:sstemp5um}. The maximum temperature rise at the center of the optical beam was 90 K.


.





\bibliography{references}